\documentclass[aps,prl,twocolumn,groupedaddress,longbibliography]{revtex4-2}

\usepackage{siunitx}

\usepackage[colorlinks,citecolor=blue,linkcolor=blue,urlcolor=blue]{hyperref}

\usepackage{amsmath}
\usepackage{mathtools}
\usepackage{esdiff}
\usepackage{amssymb}
\usepackage{graphicx, float}
\usepackage[caption=false]{subfig}
\usepackage{array}
\usepackage{mdwlist}
\usepackage{bbold}
\usepackage{stmaryrd}

\usepackage{listings} 
\usepackage{color} 
\definecolor{mygreen}{RGB}{28,172,0} 
\definecolor{mylilas}{RGB}{170,55,241}

\newcommand{\hhat}{\hat{H}}
\newcommand{\hsys}{\hat{H}_{\rm{sys}}}
\newcommand{\hbath}{\hat{H}_{\rm{bath}}}
\newcommand{\hint}{\hat{H}_{\rm{int}}}
\newcommand{\Gdiag}{\tilde{\Gamma}_{\rm{diag}}}
\newcommand{{\Goff}}{\tilde{\Gamma}_{\rm{off}}}
\newcommand{\rmd}{\mathrm{d}}
\newcommand{\rmi}{\mathrm{i}}

\newcommand{\bea}{\begin{equation}\begin{aligned}}
\newcommand{\eea}{\end{aligned}\end{equation}}

\DeclareMathOperator*{\argmin}{arg\,min}

\newcommand{\micron}{\rm{\si{\micro\meter}}}
\begin{document}

\title{Effective bath model for arrays of coupled non-Hermitian nanoresonators}

\author{Vincent Vinel}
\thanks{These two authors contributed equally.}
\author{Zejian Li}
\thanks{These two authors contributed equally.}
\author{Adrien Borne}
\author{Adrien Bensemhoun}
\author{Ivan Favero}
\author{Cristiano Ciuti}
\author{Giuseppe Leo}
\email[]{giuseppe.leo@u-paris.fr}
\affiliation{Laboratoire Matériaux et Phénomènes Quantiques, MPQ UMR 7162, Université de Paris, CNRS, 75013, Paris, France}



\begin{abstract}
Nanophotonics systems have recently been studied under the perspective of non-Hermitian physics. Given their potential for wavefront control, nonlinear optics and quantum optics, it is crucial to develop predictive tools to assist their design. We present here a simple model relying on the coupling to an effective bath consisting of a continuum of modes to describe systems of coupled resonators, and test it on dielectric nanocylinder chains accessible to experiments.  The effective coupling constants, which depend non-trivially on the distance between resonators, are extracted from numerical simulations in the case of just two coupled elements. The model predicts successfully the dispersive and reactive nature of modes for configurations with multiple resonators, as validated by numerical solutions. It can be applied to larger systems, which are hardly solvable with finite-element approaches.
\end{abstract}

\maketitle

\section{Introduction}

Nanophotonics deals with light behavior at nanoscopic scale, and encompasses plasmonics \cite{Chu2019ACS,Teperik2013_Degiron}, photonic crystals \cite{Kosaka_PRB1998_superprism, Edrington_AdvMat_2001_PolymerPhotCryst}, and metamaterial optics \cite{Decker_AdvOptMat_2015_huygens,Liu_NanoLetter_2016,Gigli_chair_2021,Gigli2019Frontier,Koshelev_review_2020}. While light confinement is a central challenge for these domains \cite{Gigli2019Frontier,Koshelev_review_2020,Koshelev288_Science_2020,Jin_Bowers_2021_HQ}, conceiving or exploring low quality factor (Q) systems is not always straightforward, as scattering or dissipative processes are omnipresent. In the past decade, nanophotonics has greatly benefited from non-Hermitian physics \cite{Lupu:13,Cortes2020_Qplasmonics}, which corresponds to the study of either time-dependent Schr{\"o}dinger equations or time-independent Schr{\"o}dinger equations with operators that are not Hermitian \cite{moiseyev2011_nonHermitian,Bender_2007_nonHermitian}. Those non-Hermitian terms describe dissipative processes such as the interaction with the environment. Non-Hermitian approaches have already been adopted in plasmonics \cite{Alaeian2014PRB,Lupu:13,Cortes2020_Qplasmonics} and more recently their dielectric counterparts \cite{Gigli2020_lalanne} have also drawn a growing interest for conceiving novel miniaturized optical devices for wavefront shaping \cite{Gigli_chair_2021,GigliIEEE_chair2020}, harmonic generation \cite{GM2019_2ndHarm,Gigli2019Frontier,Koshelev288_Science_2020} and quantum photonics \cite{Marino:19}. \\

\begin{figure}[t!]
\begin{center}
\includegraphics[width=0.46\textwidth]{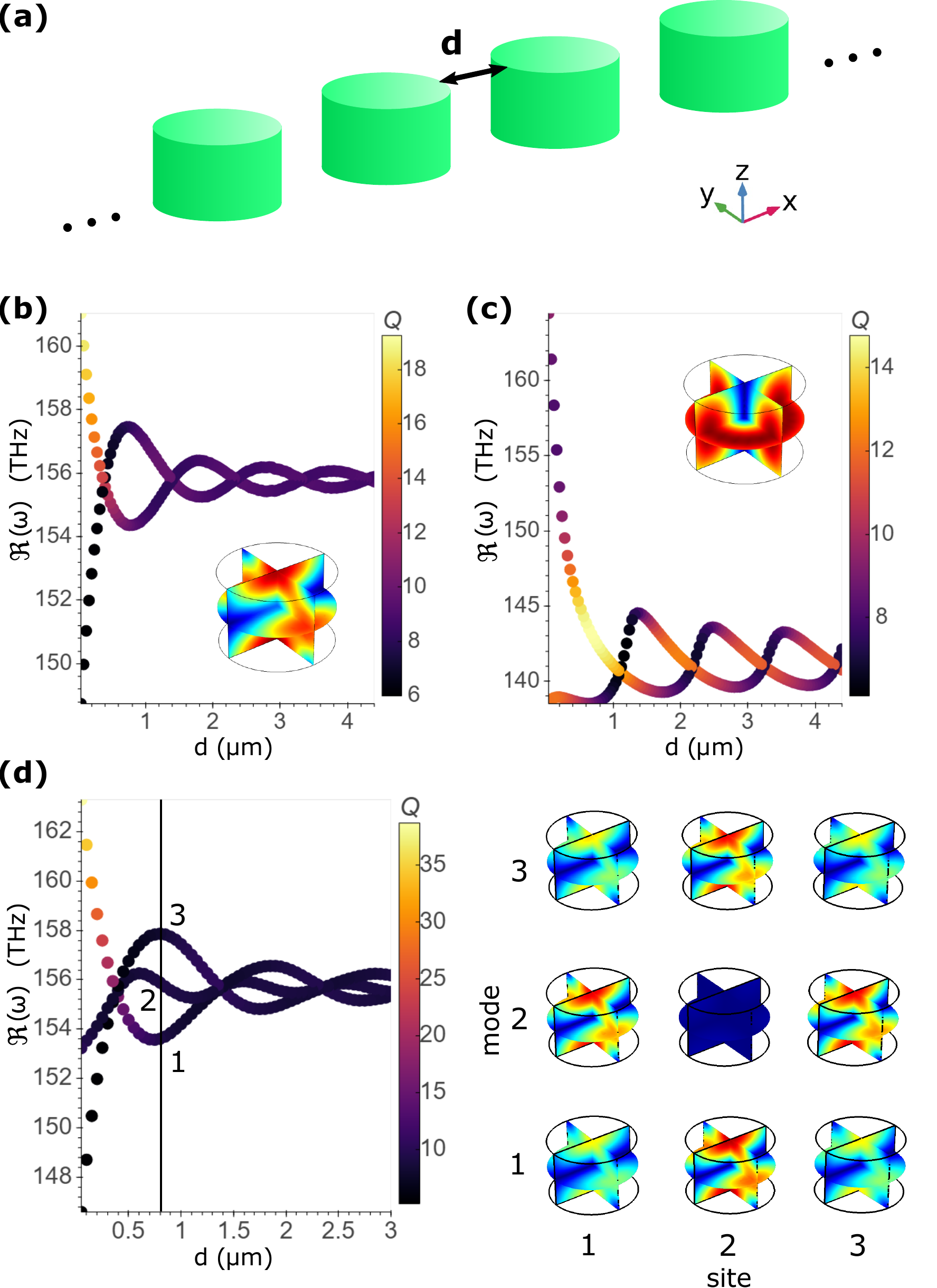}
\end{center}
\caption{\label{fig1} AlGaAs nanocylinder ensembles and numerical simulations for short chains. \textbf{(a)}: schematic of the system. The gap denotes edge-to-edge distance between nanocylinders. \textbf{(b)} and \textbf{(c)}: numerical simulation for $N=2$: real part of the eigenfrequency $\omega$ as a function of gap for MD$_{x}$ and MD$_{z}$ respectively, with colorbar providing the quality factor $Q = \Re(\omega)/2\Im(\omega)$. Inset: 3D plot of the electric field norm in a nanocylinder. \textbf{(d)}  case $N=3$ for MD$_{x}$. Electric field norm inside nanocylinders for each mode for a gap of $0.75~\micron$. Normalized powers inside the resonators for each mode are 1: (0.27, 0.47, 0.26); 2: (0.50, 0, 0.50); 3: (0.22, 0.56, 0.22).}
\end{figure}

In this work, we theoretically study one-dimensional (1D) chains of $N$ dissipative dielectric nanoresonators by using an analytical non-Hermitian quantum model. Our approach is to feed the model with complex coupling parameters obtained from a numerical simulation in the case $N= 2$, and then validate it with longer chains by comparison with brute-force calculations. It thereby becomes a predictive tool, essential for overcoming numerical limitations associated to the study and design of larger systems of interest. This model includes both direct coupling between resonators and coupling mediated by the electromagnetic continuum acting as a reservoir. Such couplings are determined by fitting the numerical solution of Maxwell's equations in the case of two coupled resonators. The latter is obtained via finite-element-method (FEM) simulations on our trial system, which consists of chains of equidistant Aluminum Gallium Arsenide (AlGaAs) nanocylinders along $x$ [Fig.~\ref{fig1}(a)]. Their size determines the spatial and spectral properties of their eigenmodes \cite{Carletti2015_OSA,Gili:16}. The numerical integration domain is bounded by a perfectly matched layer (PML) that suppresses spurious electromagnetic-field reflection at the edges of the integration domain. We focus on the hybrid modes stemming from from magnetic dipoles (MDs) of single nanocylinders (see more details in supplementary material). Their radius and height are set at 300~nm and 400~nm respectively, to lift the spectral degeneracy between out-of-plane (MD$_{z}$) and in-plane (MD$_{x}$ and MD$_{y}$) magnetic dipolar modes, the coupling strength being dictated by the gap $d$ between the nanocylinders. Figure \ref{fig1}(b,c) displays $N=2$ chain spectra for hybrid MD$_{x}$ and MD$_{z}$ as a function of the gap. A pair of bonding and antibonding modes is formed, corresponding to two complex eigenfrequencies in phase opposition. In the following, we will neglect the coupling between two MDs oriented along different directions, since the field overlap between two non-collinear MDs is 3 orders of magnitude smaller than the one between two collinear MDs. We identify two coupling regimes: an exponential-like decay of the frequency splitting for small gaps; and a pseudo-periodic variation of the complex eigenfrequencies for larger gaps. Outside each leaky resonator, the electromagnetic field decreases much slower than for high-Q systems like micropillars \cite{St-Jean2017} or whispering gallery mode resonators \cite{Armani2003,Baker:14,Parrain_Baker_OtpExp2015,Roland:20}. When more than one resonator is involved, the corresponding eigenfrequencies oscillate and result in non-zero energy splitting. A good figure of merit to assess the range of this interaction is the scattering cross-section of each resonator mode, which can be as large as ten times the cylinder cross-section in the case of MD \cite{Carletti2015_OSA,Gigli2020_lalanne}. The eigenfrequency spectra of a pair of MD eigenmodes exhibit degeneracy points which occur at different gap values for real and imaginary parts and depend on the specific mode. This feature has already been reported recently \cite{Pichugin2019} and offers interesting perspectives for meta-optics design and band engineering. Additionally, the coupling between MDs modifies the quality factor of those modes. For a single AlGaAs nanocylinder of the same dimensions, the Q factor of the MD$_{z}$ is 7 and that the MD$_{x}$ and MD$_{y}$ is 5.5: our simulations thus confirm that even long-range coupling between nanoresonators can increase Q factor, although light remains poorly confined in those structures, which justifies the introduction of a non-Hermitian formalism. \\

\section{Non-Hermitian quantum formalism}

To describe the dynamics of an ensemble of leaky resonators, let us consider an Hamiltonian consisting of a tight-binding term for the nanoresonators coupled to an effective bath.
The non-Hermitian character is obtained by tracing out the bath degrees of freedom. 
Our basic idea is to extract the effective bath parameters from exact FEM results for two coupled resonators, and then use the formalism to predict the modes for an arbitrary number and arrangement of nanoresonators.
In the rotating-wave approximation, the Hamiltonian reads (with $\hbar = 1$):

\begin{align}
        \hhat =& \hsys + \hbath + \hint,\\
        \hsys =& \sum_{j} \omega_{j} \hat{a}_{j}^{\dagger} \hat{a}_{j} +  \sum_{j,j'}J(d_{j j'}) (\hat{a}_{j}^{\dagger} \hat{a}_{j'} + \mathrm{h.c.}),\\
        \hbath =& \int\rmd\eta~ \omega_{\eta} \hat{\alpha}_{\eta}^{\dagger} \hat{\alpha}_{\eta}, \label{eq:bath}\\
        \hint =& \sum_{j}\int\rmd\eta~ \rmi(g_{j\eta} \hat{a}_{j}^{\dagger} \hat{\alpha}_{\eta} - g_{j\eta}^{*} \hat{\alpha}_{\eta}^{\dagger} \hat{a}_{j}).
\end{align}

\noindent $\hsys$ is the bare resonators' Hamiltonian, where each nanoresonator has one mode with annihilation operator $\hat{a}_{j}$ and frequency $\omega_{j}$, $j$ and $j'$ denoting different resonators. In the following, we will consider equally spaced identical resonators ($\omega_{j} = \omega_{0}$, $ d_{jj'}= d$). $J(d)$ describes the coherent coupling between two resonators via evanescent field, and is essentially determined by the distance $d$ between them for a given set of resonators. The second term $\hbath$ describes the continuum of radiation modes represented by the annihilation operators $\hat{\alpha}_{\eta}$ indexed by $\eta$, and $\hint$ is the interaction Hamiltonian between the system and the bath, where $g_{j\eta}$ is the coupling between the $j^{th}$ nanoresonator and the radiation mode $\hat{\alpha}_{\eta}$. The operators $\hat{a}_{j}$ and $\hat{\alpha}_{\eta}$ obey bosonic commutation relations, i.e. $[\hat{a}_{j}, \hat{a}^{\dagger}_{j'}] = \delta_{jj'}$ and $[\hat{\alpha}_{\eta}, \hat{\alpha}^{\dagger}_{\eta'}] = \delta(\eta-\eta')$. All commutators between $\hat{a}_j$ or $\hat{a}^\dagger_j$ and $\hat{\alpha}_\eta$ or $\hat{\alpha}^\dagger_\eta$ are taken to be zero. In the Heisenberg picture, one can derive the quantum Langevin equation \cite{Ciuti2006Input} (see Supplementary Material for the derivation) for $\hat{a}_{j}$:

\bea\label{eq_dyn}
\dfrac{\rmd \hat{a}_j}{\rmd t} = -\rmi\left[\omega_0 \hat{a}_j+J(d)\sum_{j'}\hat{a}_{j'}\right]
 -\sum_k\int_{-\infty}^\infty \rmd t' \Gamma_{jk}(t-t')\hat{a}_k(t')+\hat{F}_j(t),
\eea
where we have defined the damping kernel $\Gamma_{jk}(\tau)=\Theta(\tau)\int\rmd\eta~ g_{j\eta}g_{k\eta}^* e^{-\rmi\omega_\eta \tau}$ and the Langevin force $\hat{F}_j(t)=\int\rmd\eta~g_{j\eta}e^{-\rmi\omega_\eta(t-t_0)}\hat{\alpha}_\eta(t_0)$ where $t_0\rightarrow - \infty$.

\noindent After applying Fourier transform, adopting the convention $\tilde{A}(\omega)=\int_{-\infty}^\infty \rmd t e^{\rmi\omega t}A(t)$, to Eq.~(\ref{eq_dyn}), we obtain the equations in the frequency domain:
\bea
\omega\tilde{a}_j(\omega) = \omega_0\tilde{a}_j(\omega)+J(d)\sum_{j'}\tilde{a}_{j'}(\omega) -\rmi\sum_k\tilde{\Gamma}_{jk}(\omega)\tilde{a}_k(\omega)+\rmi\tilde{F}_j(\omega),
\eea
which can be cast into the following matrix form
\begin{equation}\label{eq:mat}
    \mathcal{M}(\omega,d)
    \begin{bmatrix}
    \tilde{a}_1(\omega) \\ \vdots \\ \tilde{a}_N(\omega)
    \end{bmatrix}
    +\rmi
    \begin{bmatrix}
    \tilde{F}_1(\omega) \\ \vdots \\ \tilde{F}_N(\omega)
    \end{bmatrix}
    =0,
\end{equation}
with eigenvalues $\lambda^{(i)}$ that can be obtained by diagonalization.
We can then solve for the resonant frequencies
\begin{equation}
    \Omega_\text{res} = \left\{\omega_{i}^\star\in \mathbb{R}~|~\exists i,~\omega_{i}^\star=\argmin_\omega |\lambda^{(i)}(\omega)|^2 \right\},
\end{equation}
which, by definition, give local maxima of the amplitude of the frequency response,
and the corresponding damping
\begin{equation}
    \gamma_i^\star = - \Im [\lambda^{(i)}(\omega_i^\star)].
\end{equation}
For a compact notation, we can assign the complex frequency to the resonant mode $ \omega^\text{res}_i = \omega^\star_i - \rmi \gamma_i^\star$. In the case $N=2$, with identical resonators, the matrix can be explicitly written as :

\begin{equation}
    \mathcal{M}(\omega, d) = 
    \begin{bmatrix}
  \omega_{0} - \omega - \rmi \Gdiag(\omega,d) &
   J(d) - \rmi \Goff(\omega,d) \\
   J(d) - \rmi \Goff(\omega,d) &
   \omega_{0} - \omega - \rmi \Gdiag(\omega,d) 
   \end{bmatrix}
   \label{matrix}
\end{equation}

\noindent where we have further assumed $\tilde{\Gamma}_{11}=\tilde{\Gamma}_{22}:=\Gdiag$ and $\tilde{\Gamma}_{12}=\tilde{\Gamma}_{21}:=\tilde{\Gamma}_\text{off}$ by symmetry. \\

The coupling functions $J$, $\Gdiag$ and $\Goff$ can be fitted from the simulation results presented in Fig.~\ref{fig1}. To simplify the treatment, we expand them to first order in $\omega$, which allows us to determine a set of possible reservoir functions from the simulation of the $N=2$ system (see supplementary material for a detailed derivation).
For an $N$-resonator chain, the matrix in Eq.~(\ref{eq:mat}) can be written as:

\begin{equation}
    \mathcal{M}_{N}(\omega, d) = (\omega - \omega_{0}) I(N) + J(d) Sec(N) - \rmi \big( \tilde{\Gamma}_{jk}(\omega,d) \big)_{\{j,k\}\in \llbracket1;N\rrbracket^2}
\end{equation}

\noindent with $I(N)$ being the $N\times N$ identity matrix, and $Sec(N)$ the $N\times N$ secondary diagonal matrix. \\

\section{Results and discussion}

\begin{figure}[htbp]
\centering
\includegraphics[width=0.46\textwidth]{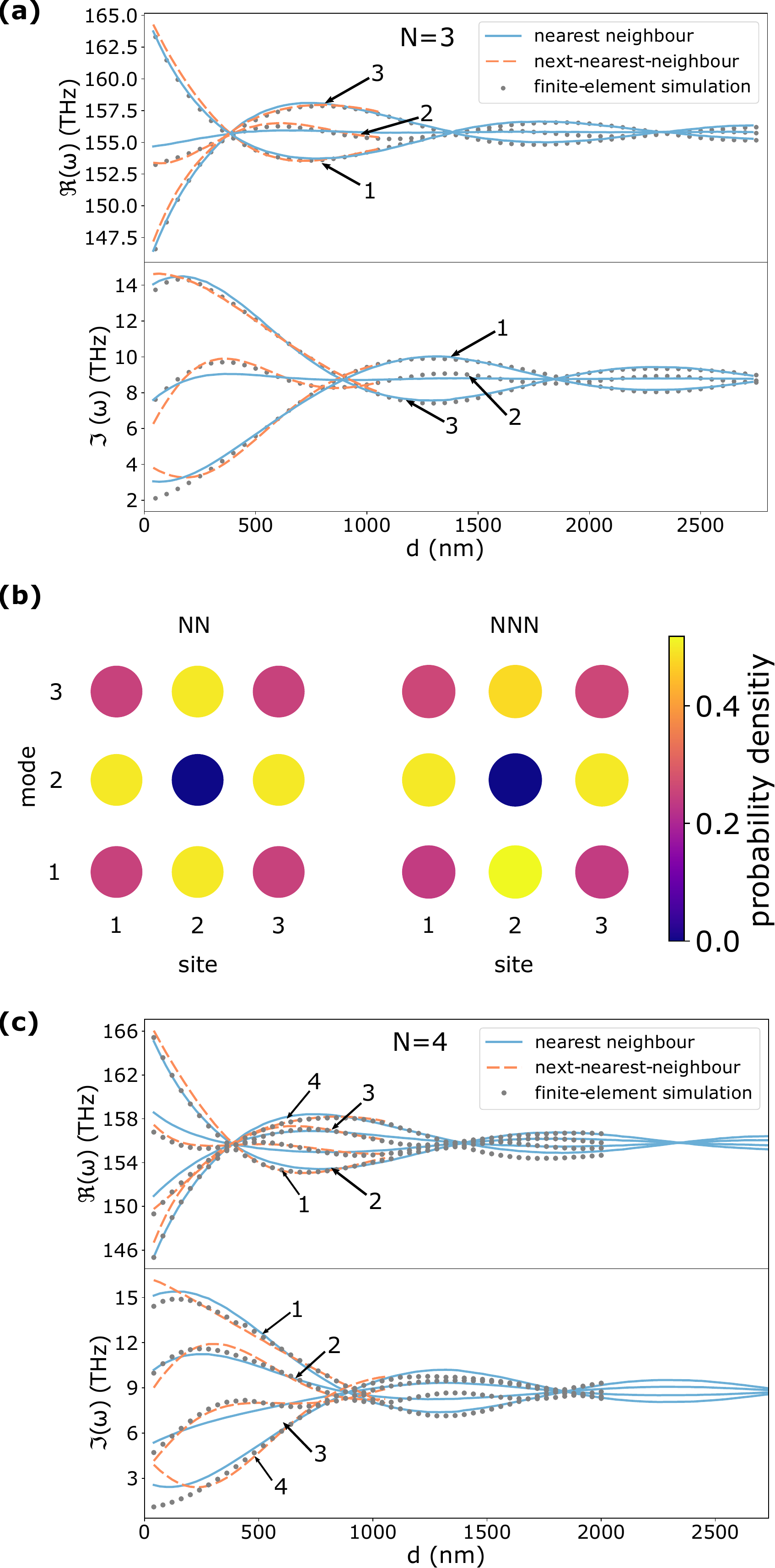}
\caption{ Comparison between the analytical model and the FEM simulations. \textbf{(a)} Case $N=3$: real part (top panel) and imaginary part (bottom panel) of the eigenfrequency $\omega$ as a function of gap for MD$_{x}$. \textbf{(b)} Probability density map in the case $N=3$, depending if next-nearest neighbor coupling is implemented (right, NNN) or not (left, NN). Modes are labelled according to Fig.~\ref{fig1}(d). \textbf{(c)} Case $N=4$: real part (top panel) and imaginary part (bottom panel) of the eigenfrequency $\omega$ as a function of gap for MD$_{x}$.}
\label{fig2}
\end{figure}

The same model is used to predict the modes in the presence of three resonators [Fig.~\ref{fig2}(a) solid blue line] and four resonators [Fig.~\ref{fig2}(c) solid blue line], using only what we learned from the $N = 2$ case behavior as a function of the gap. The main features of the fully numerical simulations are captured by our model: exponential-like decay, pseudo periodicity, and degeneracy points of both imaginary and real parts of the eigenfrequencies. All this confirms that a relatively simple analytical non-Hermitian formalism can predict the physics of non-trivial nanophotonic systems. \\

Strikingly, degeneracy points for real parts on one hand, and for imaginary parts on the other hand, arise for the same gaps in the cases $N=2$, $3$ and $4$. The fact that they are independent of $N$ indicates that they only depend on single resonator's characteristics. The numerical spectra of the eigenmodes in the cases $N=3$ and $4$ are fairly described by our formalism (see the relative error in the Fig. S4 of the supplementary material). Furthermore, following the correspondence between the probability density and the power stored inside each resonator, the eigenmodes solution of the non-unitary dynamics in the case $N=3$ [Fig.~\ref{fig2}(b)] displays more symmetric results than the FEM simulations [Fig.~\ref{fig1}(d)]. Therefore, in order to refine our model, we explore coupling to the next-nearest neighbor.\\

It is important to note that the coupling between two resonators $j$ and $j+2$ should be different whether a resonator $j+1$ is present or not. This implies that next-nearest-neighbor coupling cannot be extracted from the $N=2$ case, but from the $N=3$ at least. Therefore, the knowledge of long-range coupling in a chain of a $N$ nanoresonators depends on the knowledge of the corresponding $N-1$ chain. However, the resolution of such a model would prove tedious, with had-to-extract coupling constants through iterative calculation processes, resulting in numerical challenges to predict modes of longer chains. From the information extracted in the case $N=2$, we performed an analytical calculation of the $N=3$ (resp. N=4) chain with simplified next-nearest-neighbor coupling [Fig.~\ref{fig2}(a) and (b)] (resp. (c)) to clarify whether it could improve our model. For this purpose, we introduce $J(2r+2d) + \rmi\Goff (2r+2d)$, the coupling between two nanoresonators separated by $2r+2d$ , where r is the radius of the nanocylinder. Conservative and dissipative coupling constants, $J(2r+2d)$ and $\Goff (2r+2d)$, were extracted from Maxwell's equations solutions in the case $N=2$. This simplified next-nearest-neighbor coupling improves the agreement of probability densities [Fig.~\ref{fig2}(b)] and spectra [Fig.~\ref{fig2}(a) and (c)] with FEM simulations (see the relative error in the Fig. S5 of the supplementary material). Such a refinement confirms that those modes involve long-range interaction between coupled resonators. \\

\begin{figure}
\begin{center}
\includegraphics[width=0.46\textwidth]{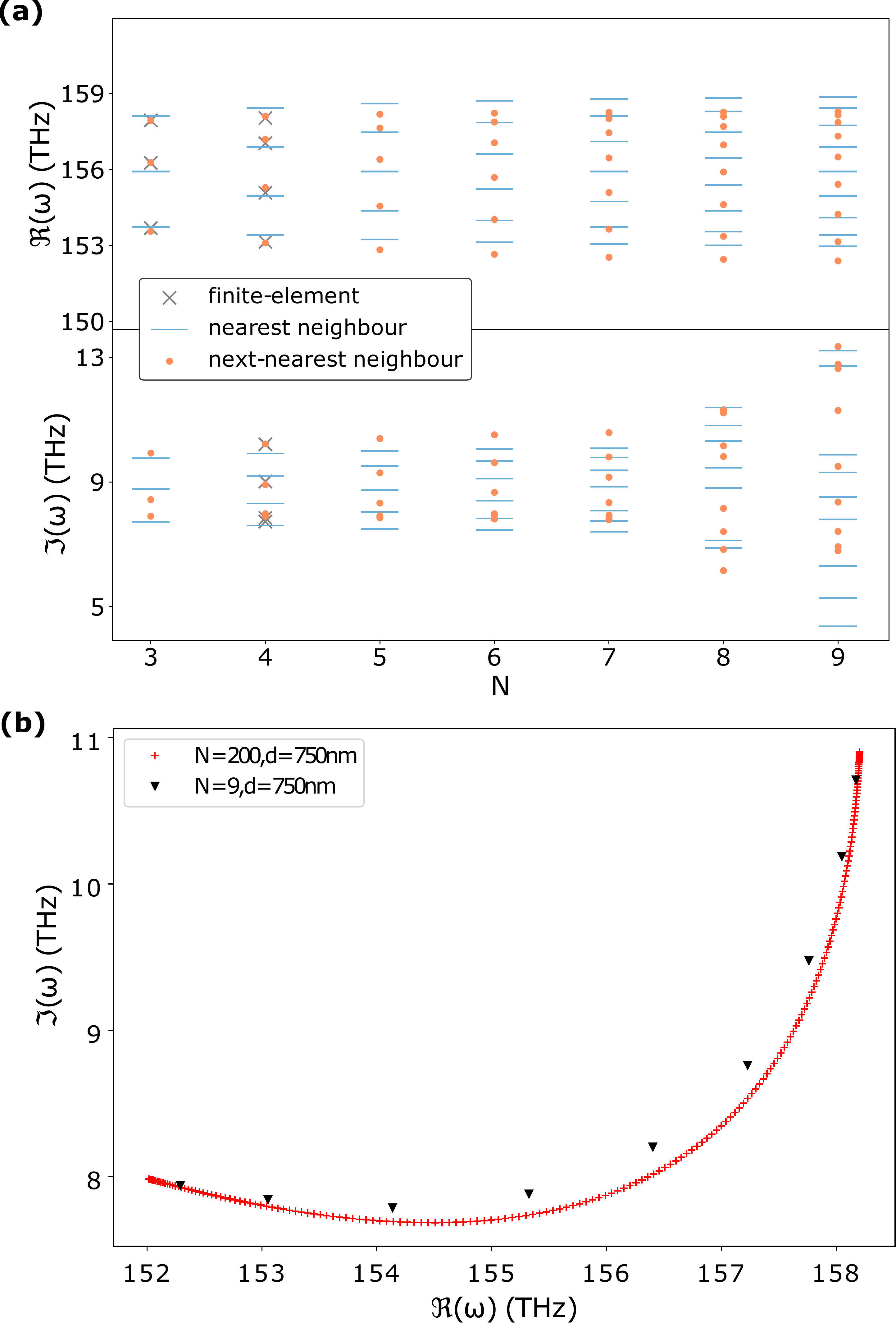}
\caption{Comparison between numerical simulations and analytical model for $d=750~$nm. \textbf{(a)} Real part (top panel) and imaginary part (bottom panel) of the eigenfrequencies $\omega$ vs. the number N of resonators in the chain. Blue dashes: spectra calculated with the Hamiltonian model; grey crosses: FEM simulations. \textbf{(b)} Complex-plane representation of the analytically calculated eigenfrequencies for chains of $N=9$ (black triangles) and $N=200$ (red crosses).}
\label{fig3}
\end{center}
\end{figure}

From Fig.~\ref{fig2}, it also appears that the agreement between the analytical non-Hermitian model and the numerical Maxwell's equations resolution is more significant for larger gaps. This can be ascribed to the hypothesis of linear dependence of the coupling constants on $\omega$, which enabled us to fit the complex eigenfrequencies. For a narrower gap between nanocylinders, the field overlap grows stronger, which implies a wider frequency splitting of the eigenmodes \cite{Zhang_2012_coupling,Vial_2016_coupling}, as well as a deformation of near-field both outside and inside the resonators. In this regime, the coupling can no longer be expressed as a perturbation, increasing the deviation of our analytical model from brute-force calculations. Finally, we computed spectra for 1D equidistant chains with different number of sites $N$ (Fig.~\ref{fig3}). For a chain of $N$ resonators, $N$ hybrid modes are expected for a given MD. While analytical non-Hermitian calculations derive dispersion from eigenmodes of the $N$-site chains, numerical simulations are quickly limited by the size of the calculation space, which is defined by the PML size and the meshing of the system. Indeed, for $N=6$, FEM simulations require about 100 GBytes of RAM, which meets the upper limit of our local calculation resources. Additionally, hybridization of the eigenmodes with the PML modes prevents us to identify them properly. This is less prominent for MD$_{y}$ and MD$_{z}$ hybrid modes, whose symmetries differ more from spherical PML than MD$_{x}$ modes. This observation underpins the necessity of finding alternative means to model and design coupled nanophotonic systems of larger size. In opposition, analytical resolution of the tight-binding non-Hermitian problem is possible for large $N$, as illustrated in Fig.~\ref{fig3}(b) for $N=200$, where we assumed no frequency dependence in the coupling constants for this specific calculation. In this case, the eigenfrequencies of the 1D chain tend to form a continuum of modes, whose spectral localization depends on the coupling constants of the system. \\

\section{Conclusion}

Overall, the analytical non-Hermitian quantum formalism used here offers adequate means to compute modal information on 1D chains of $N$ resonators, and it can be naturally extended to nonlinear quantum optics \cite{Li2021}. By increasing the length of the chain, one could study the transition between a discrete ensemble of nanoresonators and a photonic crystal. Considering the wide variety of geometries and properties accessible via nanostructured dielectric materials, we envisage that such nanophotonic systems could be a promising model for implementing more complex interactions in 1D, 2D or 3D metasystems. They could also constitute toy systems to study for example topological edge states \cite{Kivshar_TopoRev_2020,Kruk_Nature_2019}. A direct application for such a predictive model could be the improvement of Q factor optimization of leaky nanoresonators through non-Hermitian coupling.  
 \\

\begin{acknowledgments}
This work has been supported by the ANR grants NANOPAIR (ANR-18-CE92-0043) and NOMOS (ANR-18-CE24-0026).
The authors would like to thank Carlo Gigli his contribution to the numerical simulations platform. Data underlying the results presented in this paper are not publicly available at this time but may be obtained from the authors upon reasonable request. The authors declare no conflicts of interest.
\end{acknowledgments}

\bibliography{Main_arXiv.bib}

\begin{thebibliography}{34}%
\makeatletter
\providecommand \@ifxundefined [1]{%
 \@ifx{#1\undefined}
}%
\providecommand \@ifnum [1]{%
 \ifnum #1\expandafter \@firstoftwo
 \else \expandafter \@secondoftwo
 \fi
}%
\providecommand \@ifx [1]{%
 \ifx #1\expandafter \@firstoftwo
 \else \expandafter \@secondoftwo
 \fi
}%
\providecommand \natexlab [1]{#1}%
\providecommand \enquote  [1]{``#1''}%
\providecommand \bibnamefont  [1]{#1}%
\providecommand \bibfnamefont [1]{#1}%
\providecommand \citenamefont [1]{#1}%
\providecommand \href@noop [0]{\@secondoftwo}%
\providecommand \href [0]{\begingroup \@sanitize@url \@href}%
\providecommand \@href[1]{\@@startlink{#1}\@@href}%
\providecommand \@@href[1]{\endgroup#1\@@endlink}%
\providecommand \@sanitize@url [0]{\catcode `\\12\catcode `\$12\catcode
  `\&12\catcode `\#12\catcode `\^12\catcode `\_12\catcode `\%12\relax}%
\providecommand \@@startlink[1]{}%
\providecommand \@@endlink[0]{}%
\providecommand \url  [0]{\begingroup\@sanitize@url \@url }%
\providecommand \@url [1]{\endgroup\@href {#1}{\urlprefix }}%
\providecommand \urlprefix  [0]{URL }%
\providecommand \Eprint [0]{\href }%
\providecommand \doibase [0]{https://doi.org/}%
\providecommand \selectlanguage [0]{\@gobble}%
\providecommand \bibinfo  [0]{\@secondoftwo}%
\providecommand \bibfield  [0]{\@secondoftwo}%
\providecommand \translation [1]{[#1]}%
\providecommand \BibitemOpen [0]{}%
\providecommand \bibitemStop [0]{}%
\providecommand \bibitemNoStop [0]{.\EOS\space}%
\providecommand \EOS [0]{\spacefactor3000\relax}%
\providecommand \BibitemShut  [1]{\csname bibitem#1\endcsname}%
\let\auto@bib@innerbib\@empty
\bibitem [{\citenamefont {Chu}\ \emph {et~al.}(2019)\citenamefont {Chu},
  \citenamefont {Gr\'eboval}, \citenamefont {Goubet}, \citenamefont {Martinez},
  \citenamefont {Livache}, \citenamefont {Qu}, \citenamefont {Rastogi},
  \citenamefont {Bresciani}, \citenamefont {Prado}, \citenamefont {Suffit},
  \citenamefont {Ithurria}, \citenamefont {Vincent},\ and\ \citenamefont
  {Lhuillier}}]{Chu2019ACS}%
  \BibitemOpen
  \bibfield  {author} {\bibinfo {author} {\bibfnamefont {A.}~\bibnamefont
  {Chu}}, \bibinfo {author} {\bibfnamefont {C.}~\bibnamefont {Gr\'eboval}},
  \bibinfo {author} {\bibfnamefont {N.}~\bibnamefont {Goubet}}, \bibinfo
  {author} {\bibfnamefont {B.}~\bibnamefont {Martinez}}, \bibinfo {author}
  {\bibfnamefont {C.}~\bibnamefont {Livache}}, \bibinfo {author} {\bibfnamefont
  {J.}~\bibnamefont {Qu}}, \bibinfo {author} {\bibfnamefont {P.}~\bibnamefont
  {Rastogi}}, \bibinfo {author} {\bibfnamefont {F.~A.}\ \bibnamefont
  {Bresciani}}, \bibinfo {author} {\bibfnamefont {Y.}~\bibnamefont {Prado}},
  \bibinfo {author} {\bibfnamefont {S.}~\bibnamefont {Suffit}}, \bibinfo
  {author} {\bibfnamefont {S.}~\bibnamefont {Ithurria}}, \bibinfo {author}
  {\bibfnamefont {G.}~\bibnamefont {Vincent}},\ and\ \bibinfo {author}
  {\bibfnamefont {E.}~\bibnamefont {Lhuillier}},\ }\href
  {https://doi.org/10.1021/acsphotonics.9b01015} {\bibfield  {journal}
  {\bibinfo  {journal} {ACS Photonics}\ }\textbf {\bibinfo {volume} {6}},\
  \bibinfo {pages} {2553} (\bibinfo {year} {2019})},\ \Eprint
  {https://arxiv.org/abs/https://doi.org/10.1021/acsphotonics.9b01015}
  {https://doi.org/10.1021/acsphotonics.9b01015} \BibitemShut {NoStop}%
\bibitem [{\citenamefont {Teperik}\ and\ \citenamefont
  {Degiron}(2012)}]{Teperik2013_Degiron}%
  \BibitemOpen
  \bibfield  {author} {\bibinfo {author} {\bibfnamefont {T.~V.}\ \bibnamefont
  {Teperik}}\ and\ \bibinfo {author} {\bibfnamefont {A.}~\bibnamefont
  {Degiron}},\ }\href {https://doi.org/10.1103/PhysRevLett.108.147401}
  {\bibfield  {journal} {\bibinfo  {journal} {Phys. Rev. Lett.}\ }\textbf
  {\bibinfo {volume} {108}},\ \bibinfo {pages} {147401} (\bibinfo {year}
  {2012})}\BibitemShut {NoStop}%
\bibitem [{\citenamefont {Kosaka}\ \emph {et~al.}(1998)\citenamefont {Kosaka},
  \citenamefont {Kawashima}, \citenamefont {Tomita}, \citenamefont {Notomi},
  \citenamefont {Tamamura}, \citenamefont {Sato},\ and\ \citenamefont
  {Kawakami}}]{Kosaka_PRB1998_superprism}%
  \BibitemOpen
  \bibfield  {author} {\bibinfo {author} {\bibfnamefont {H.}~\bibnamefont
  {Kosaka}}, \bibinfo {author} {\bibfnamefont {T.}~\bibnamefont {Kawashima}},
  \bibinfo {author} {\bibfnamefont {A.}~\bibnamefont {Tomita}}, \bibinfo
  {author} {\bibfnamefont {M.}~\bibnamefont {Notomi}}, \bibinfo {author}
  {\bibfnamefont {T.}~\bibnamefont {Tamamura}}, \bibinfo {author}
  {\bibfnamefont {T.}~\bibnamefont {Sato}},\ and\ \bibinfo {author}
  {\bibfnamefont {S.}~\bibnamefont {Kawakami}},\ }\href
  {https://doi.org/10.1103/PhysRevB.58.R10096} {\bibfield  {journal} {\bibinfo
  {journal} {Phys. Rev. B}\ }\textbf {\bibinfo {volume} {58}},\ \bibinfo
  {pages} {R10096} (\bibinfo {year} {1998})}\BibitemShut {NoStop}%
\bibitem [{\citenamefont {Edrington}\ \emph {et~al.}(2001)\citenamefont
  {Edrington}, \citenamefont {Urbas}, \citenamefont {DeRege}, \citenamefont
  {Chen}, \citenamefont {Swager}, \citenamefont {Hadjichristidis},
  \citenamefont {Xenidou}, \citenamefont {Fetters}, \citenamefont
  {Joannopoulos}, \citenamefont {Fink},\ and\ \citenamefont
  {Thomas}}]{Edrington_AdvMat_2001_PolymerPhotCryst}%
  \BibitemOpen
  \bibfield  {author} {\bibinfo {author} {\bibfnamefont {A.~C.}\ \bibnamefont
  {Edrington}}, \bibinfo {author} {\bibfnamefont {A.~M.}\ \bibnamefont
  {Urbas}}, \bibinfo {author} {\bibfnamefont {P.}~\bibnamefont {DeRege}},
  \bibinfo {author} {\bibfnamefont {C.~X.}\ \bibnamefont {Chen}}, \bibinfo
  {author} {\bibfnamefont {T.~M.}\ \bibnamefont {Swager}}, \bibinfo {author}
  {\bibfnamefont {N.}~\bibnamefont {Hadjichristidis}}, \bibinfo {author}
  {\bibfnamefont {M.}~\bibnamefont {Xenidou}}, \bibinfo {author} {\bibfnamefont
  {L.~J.}\ \bibnamefont {Fetters}}, \bibinfo {author} {\bibfnamefont {J.~D.}\
  \bibnamefont {Joannopoulos}}, \bibinfo {author} {\bibfnamefont
  {Y.}~\bibnamefont {Fink}},\ and\ \bibinfo {author} {\bibfnamefont {E.~L.}\
  \bibnamefont {Thomas}},\ }\href
  {https://doi.org/https://doi.org/10.1002/1521-4095(200103)13:6<421::AID-ADMA421>3.0.CO;2-\#}
  {\bibfield  {journal} {\bibinfo  {journal} {Advanced Materials}\ }\textbf
  {\bibinfo {volume} {13}},\ \bibinfo {pages} {421} (\bibinfo {year}
  {2001})}\BibitemShut {NoStop}%
\bibitem [{\citenamefont {Decker}\ \emph {et~al.}(2015)\citenamefont {Decker},
  \citenamefont {Staude}, \citenamefont {Falkner}, \citenamefont {Dominguez},
  \citenamefont {Neshev}, \citenamefont {Brener}, \citenamefont {Pertsch},\
  and\ \citenamefont {Kivshar}}]{Decker_AdvOptMat_2015_huygens}%
  \BibitemOpen
  \bibfield  {author} {\bibinfo {author} {\bibfnamefont {M.}~\bibnamefont
  {Decker}}, \bibinfo {author} {\bibfnamefont {I.}~\bibnamefont {Staude}},
  \bibinfo {author} {\bibfnamefont {M.}~\bibnamefont {Falkner}}, \bibinfo
  {author} {\bibfnamefont {J.}~\bibnamefont {Dominguez}}, \bibinfo {author}
  {\bibfnamefont {D.~N.}\ \bibnamefont {Neshev}}, \bibinfo {author}
  {\bibfnamefont {I.}~\bibnamefont {Brener}}, \bibinfo {author} {\bibfnamefont
  {T.}~\bibnamefont {Pertsch}},\ and\ \bibinfo {author} {\bibfnamefont {Y.~S.}\
  \bibnamefont {Kivshar}},\ }\href {https://doi.org/10.1002/adom.201400584}
  {\bibfield  {journal} {\bibinfo  {journal} {Advanced Optical Materials}\
  }\textbf {\bibinfo {volume} {3}},\ \bibinfo {pages} {813} (\bibinfo {year}
  {2015})}\BibitemShut {NoStop}%
\bibitem [{\citenamefont {Liu}\ \emph {et~al.}(2016)\citenamefont {Liu},
  \citenamefont {Sinclair}, \citenamefont {Saravi}, \citenamefont {Keeler},
  \citenamefont {Yang}, \citenamefont {Reno}, \citenamefont {Peake},
  \citenamefont {Setzpfandt}, \citenamefont {Staude}, \citenamefont {Pertsch},\
  and\ \citenamefont {Brener}}]{Liu_NanoLetter_2016}%
  \BibitemOpen
  \bibfield  {author} {\bibinfo {author} {\bibfnamefont {S.}~\bibnamefont
  {Liu}}, \bibinfo {author} {\bibfnamefont {M.~B.}\ \bibnamefont {Sinclair}},
  \bibinfo {author} {\bibfnamefont {S.}~\bibnamefont {Saravi}}, \bibinfo
  {author} {\bibfnamefont {G.~A.}\ \bibnamefont {Keeler}}, \bibinfo {author}
  {\bibfnamefont {Y.}~\bibnamefont {Yang}}, \bibinfo {author} {\bibfnamefont
  {J.}~\bibnamefont {Reno}}, \bibinfo {author} {\bibfnamefont {G.~M.}\
  \bibnamefont {Peake}}, \bibinfo {author} {\bibfnamefont {F.}~\bibnamefont
  {Setzpfandt}}, \bibinfo {author} {\bibfnamefont {I.}~\bibnamefont {Staude}},
  \bibinfo {author} {\bibfnamefont {T.}~\bibnamefont {Pertsch}},\ and\ \bibinfo
  {author} {\bibfnamefont {I.}~\bibnamefont {Brener}},\ }\href
  {https://doi.org/10.1021/acs.nanolett.6b01816} {\bibfield  {journal}
  {\bibinfo  {journal} {Nano Letters}\ }\textbf {\bibinfo {volume} {16}},\
  \bibinfo {pages} {5426} (\bibinfo {year} {2016})},\ \Eprint
  {https://arxiv.org/abs/https://doi.org/10.1021/acs.nanolett.6b01816}
  {https://doi.org/10.1021/acs.nanolett.6b01816} \BibitemShut {NoStop}%
\bibitem [{\citenamefont {Gigli}\ \emph {et~al.}(2021)\citenamefont {Gigli},
  \citenamefont {Marino}, \citenamefont {Artioli}, \citenamefont {Rocco},
  \citenamefont {Angelis}, \citenamefont {Claudon}, \citenamefont
  {G\'{e}rard},\ and\ \citenamefont {Leo}}]{Gigli_chair_2021}%
  \BibitemOpen
  \bibfield  {author} {\bibinfo {author} {\bibfnamefont {C.}~\bibnamefont
  {Gigli}}, \bibinfo {author} {\bibfnamefont {G.}~\bibnamefont {Marino}},
  \bibinfo {author} {\bibfnamefont {A.}~\bibnamefont {Artioli}}, \bibinfo
  {author} {\bibfnamefont {D.}~\bibnamefont {Rocco}}, \bibinfo {author}
  {\bibfnamefont {C.~D.}\ \bibnamefont {Angelis}}, \bibinfo {author}
  {\bibfnamefont {J.}~\bibnamefont {Claudon}}, \bibinfo {author} {\bibfnamefont
  {J.-M.}\ \bibnamefont {G\'{e}rard}},\ and\ \bibinfo {author} {\bibfnamefont
  {G.}~\bibnamefont {Leo}},\ }\href {https://doi.org/10.1364/OPTICA.413329}
  {\bibfield  {journal} {\bibinfo  {journal} {Optica}\ }\textbf {\bibinfo
  {volume} {8}},\ \bibinfo {pages} {269} (\bibinfo {year} {2021})}\BibitemShut
  {NoStop}%
\bibitem [{\citenamefont {Gigli}\ \emph {et~al.}(2019)\citenamefont {Gigli},
  \citenamefont {Marino}, \citenamefont {Borne}, \citenamefont {Lalanne},\ and\
  \citenamefont {Leo}}]{Gigli2019Frontier}%
  \BibitemOpen
  \bibfield  {author} {\bibinfo {author} {\bibfnamefont {C.}~\bibnamefont
  {Gigli}}, \bibinfo {author} {\bibfnamefont {G.}~\bibnamefont {Marino}},
  \bibinfo {author} {\bibfnamefont {A.}~\bibnamefont {Borne}}, \bibinfo
  {author} {\bibfnamefont {P.}~\bibnamefont {Lalanne}},\ and\ \bibinfo {author}
  {\bibfnamefont {G.}~\bibnamefont {Leo}},\ }\href
  {https://doi.org/10.3389/fphy.2019.00221} {\bibfield  {journal} {\bibinfo
  {journal} {Frontiers in Physics}\ }\textbf {\bibinfo {volume} {7}},\ \bibinfo
  {pages} {221} (\bibinfo {year} {2019})}\BibitemShut {NoStop}%
\bibitem [{\citenamefont {Koshelev}\ and\ \citenamefont
  {Kivshar}(2021)}]{Koshelev_review_2020}%
  \BibitemOpen
  \bibfield  {author} {\bibinfo {author} {\bibfnamefont {K.}~\bibnamefont
  {Koshelev}}\ and\ \bibinfo {author} {\bibfnamefont {Y.}~\bibnamefont
  {Kivshar}},\ }\href {https://doi.org/10.1021/acsphotonics.0c01315} {\bibfield
   {journal} {\bibinfo  {journal} {ACS Photonics}\ }\textbf {\bibinfo {volume}
  {8}},\ \bibinfo {pages} {102} (\bibinfo {year} {2021})},\ \Eprint
  {https://arxiv.org/abs/https://doi.org/10.1021/acsphotonics.0c01315}
  {https://doi.org/10.1021/acsphotonics.0c01315} \BibitemShut {NoStop}%
\bibitem [{\citenamefont {Koshelev}\ \emph {et~al.}(2020)\citenamefont
  {Koshelev}, \citenamefont {Kruk}, \citenamefont {Melik-Gaykazyan},
  \citenamefont {Choi}, \citenamefont {Bogdanov}, \citenamefont {Park},\ and\
  \citenamefont {Kivshar}}]{Koshelev288_Science_2020}%
  \BibitemOpen
  \bibfield  {author} {\bibinfo {author} {\bibfnamefont {K.}~\bibnamefont
  {Koshelev}}, \bibinfo {author} {\bibfnamefont {S.}~\bibnamefont {Kruk}},
  \bibinfo {author} {\bibfnamefont {E.}~\bibnamefont {Melik-Gaykazyan}},
  \bibinfo {author} {\bibfnamefont {J.-H.}\ \bibnamefont {Choi}}, \bibinfo
  {author} {\bibfnamefont {A.}~\bibnamefont {Bogdanov}}, \bibinfo {author}
  {\bibfnamefont {H.-G.}\ \bibnamefont {Park}},\ and\ \bibinfo {author}
  {\bibfnamefont {Y.}~\bibnamefont {Kivshar}},\ }\href
  {https://doi.org/10.1126/science.aaz3985} {\bibfield  {journal} {\bibinfo
  {journal} {Science}\ }\textbf {\bibinfo {volume} {367}},\ \bibinfo {pages}
  {288} (\bibinfo {year} {2020})},\ \Eprint
  {https://arxiv.org/abs/https://science.sciencemag.org/content/367/6475/288.full.pdf}
  {https://science.sciencemag.org/content/367/6475/288.full.pdf} \BibitemShut
  {NoStop}%
\bibitem [{\citenamefont {Jin}\ \emph {et~al.}(2021)\citenamefont {Jin},
  \citenamefont {Yang}, \citenamefont {Chang}, \citenamefont {Shen},
  \citenamefont {Wang}, \citenamefont {Leal}, \citenamefont {Wu}, \citenamefont
  {Gao}, \citenamefont {Feshali}, \citenamefont {Paniccia}, \citenamefont
  {Vahala},\ and\ \citenamefont {Bowers}}]{Jin_Bowers_2021_HQ}%
  \BibitemOpen
  \bibfield  {author} {\bibinfo {author} {\bibfnamefont {W.}~\bibnamefont
  {Jin}}, \bibinfo {author} {\bibfnamefont {Q.-F.}\ \bibnamefont {Yang}},
  \bibinfo {author} {\bibfnamefont {L.}~\bibnamefont {Chang}}, \bibinfo
  {author} {\bibfnamefont {B.}~\bibnamefont {Shen}}, \bibinfo {author}
  {\bibfnamefont {H.}~\bibnamefont {Wang}}, \bibinfo {author} {\bibfnamefont
  {M.~A.}\ \bibnamefont {Leal}}, \bibinfo {author} {\bibfnamefont
  {L.}~\bibnamefont {Wu}}, \bibinfo {author} {\bibfnamefont {M.}~\bibnamefont
  {Gao}}, \bibinfo {author} {\bibfnamefont {A.}~\bibnamefont {Feshali}},
  \bibinfo {author} {\bibfnamefont {M.}~\bibnamefont {Paniccia}}, \bibinfo
  {author} {\bibfnamefont {K.~J.}\ \bibnamefont {Vahala}},\ and\ \bibinfo
  {author} {\bibfnamefont {J.~E.}\ \bibnamefont {Bowers}},\ }\href
  {https://doi.org/10.1038/s41566-021-00761-7} {\bibfield  {journal} {\bibinfo
  {journal} {Nature Photonics}\ }\textbf {\bibinfo {volume} {15}},\ \bibinfo
  {pages} {346} (\bibinfo {year} {2021})}\BibitemShut {NoStop}%
\bibitem [{\citenamefont {Lupu}\ \emph {et~al.}(2013)\citenamefont {Lupu},
  \citenamefont {Benisty},\ and\ \citenamefont {Degiron}}]{Lupu:13}%
  \BibitemOpen
  \bibfield  {author} {\bibinfo {author} {\bibfnamefont {A.}~\bibnamefont
  {Lupu}}, \bibinfo {author} {\bibfnamefont {H.}~\bibnamefont {Benisty}},\ and\
  \bibinfo {author} {\bibfnamefont {A.}~\bibnamefont {Degiron}},\ }\href
  {https://doi.org/10.1364/OE.21.021651} {\bibfield  {journal} {\bibinfo
  {journal} {Opt. Express}\ }\textbf {\bibinfo {volume} {21}},\ \bibinfo
  {pages} {21651} (\bibinfo {year} {2013})}\BibitemShut {NoStop}%
\bibitem [{\citenamefont {Cortes}\ \emph {et~al.}(2020)\citenamefont {Cortes},
  \citenamefont {Otten},\ and\ \citenamefont {Gray}}]{Cortes2020_Qplasmonics}%
  \BibitemOpen
  \bibfield  {author} {\bibinfo {author} {\bibfnamefont {C.~L.}\ \bibnamefont
  {Cortes}}, \bibinfo {author} {\bibfnamefont {M.}~\bibnamefont {Otten}},\ and\
  \bibinfo {author} {\bibfnamefont {S.~K.}\ \bibnamefont {Gray}},\ }\href
  {https://doi.org/10.1063/1.5131762} {\bibfield  {journal} {\bibinfo
  {journal} {The Journal of Chemical Physics}\ }\textbf {\bibinfo {volume}
  {152}},\ \bibinfo {pages} {084105} (\bibinfo {year} {2020})},\ \Eprint
  {https://arxiv.org/abs/https://doi.org/10.1063/1.5131762}
  {https://doi.org/10.1063/1.5131762} \BibitemShut {NoStop}%
\bibitem [{\citenamefont {Moiseyev}(2011)}]{moiseyev2011_nonHermitian}%
  \BibitemOpen
  \bibfield  {author} {\bibinfo {author} {\bibfnamefont {N.}~\bibnamefont
  {Moiseyev}},\ }\href@noop {} {\emph {\bibinfo {title} {Non-Hermitian quantum
  mechanics}}}\ (\bibinfo  {publisher} {Cambridge University},\ \bibinfo {year}
  {2011})\BibitemShut {NoStop}%
\bibitem [{\citenamefont {Bender}(2007)}]{Bender_2007_nonHermitian}%
  \BibitemOpen
  \bibfield  {author} {\bibinfo {author} {\bibfnamefont {C.~M.}\ \bibnamefont
  {Bender}},\ }\href {https://doi.org/10.1088/0034-4885/70/6/r03} {\bibfield
  {journal} {\bibinfo  {journal} {Reports on Progress in Physics}\ }\textbf
  {\bibinfo {volume} {70}},\ \bibinfo {pages} {947} (\bibinfo {year}
  {2007})}\BibitemShut {NoStop}%
\bibitem [{\citenamefont {Alaeian}\ and\ \citenamefont
  {Dionne}(2014)}]{Alaeian2014PRB}%
  \BibitemOpen
  \bibfield  {author} {\bibinfo {author} {\bibfnamefont {H.}~\bibnamefont
  {Alaeian}}\ and\ \bibinfo {author} {\bibfnamefont {J.~A.}\ \bibnamefont
  {Dionne}},\ }\href {https://doi.org/10.1103/PhysRevB.89.075136} {\bibfield
  {journal} {\bibinfo  {journal} {Phys. Rev. B}\ }\textbf {\bibinfo {volume}
  {89}},\ \bibinfo {pages} {075136} (\bibinfo {year} {2014})}\BibitemShut
  {NoStop}%
\bibitem [{\citenamefont {Gigli}\ \emph {et~al.}(2020)\citenamefont {Gigli},
  \citenamefont {Wu}, \citenamefont {Marino}, \citenamefont {Borne},
  \citenamefont {Leo},\ and\ \citenamefont {Lalanne}}]{Gigli2020_lalanne}%
  \BibitemOpen
  \bibfield  {author} {\bibinfo {author} {\bibfnamefont {C.}~\bibnamefont
  {Gigli}}, \bibinfo {author} {\bibfnamefont {T.}~\bibnamefont {Wu}}, \bibinfo
  {author} {\bibfnamefont {G.}~\bibnamefont {Marino}}, \bibinfo {author}
  {\bibfnamefont {A.}~\bibnamefont {Borne}}, \bibinfo {author} {\bibfnamefont
  {G.}~\bibnamefont {Leo}},\ and\ \bibinfo {author} {\bibfnamefont
  {P.}~\bibnamefont {Lalanne}},\ }\href
  {https://doi.org/10.1021/acsphotonics.0c00014} {\bibfield  {journal}
  {\bibinfo  {journal} {ACS Photonics}\ }\textbf {\bibinfo {volume} {7}},\
  \bibinfo {pages} {1197} (\bibinfo {year} {2020})},\ \Eprint
  {https://arxiv.org/abs/https://doi.org/10.1021/acsphotonics.0c00014}
  {https://doi.org/10.1021/acsphotonics.0c00014} \BibitemShut {NoStop}%
\bibitem [{\citenamefont {{Rocco}}\ \emph {et~al.}(2020)\citenamefont
  {{Rocco}}, \citenamefont {{Gigli}}, \citenamefont {{Carletti}}, \citenamefont
  {{Marino}}, \citenamefont {{Vincenti}}, \citenamefont {{Leo}},\ and\
  \citenamefont {{De Angelis}}}]{GigliIEEE_chair2020}%
  \BibitemOpen
  \bibfield  {author} {\bibinfo {author} {\bibfnamefont {D.}~\bibnamefont
  {{Rocco}}}, \bibinfo {author} {\bibfnamefont {C.}~\bibnamefont {{Gigli}}},
  \bibinfo {author} {\bibfnamefont {L.}~\bibnamefont {{Carletti}}}, \bibinfo
  {author} {\bibfnamefont {G.}~\bibnamefont {{Marino}}}, \bibinfo {author}
  {\bibfnamefont {M.~A.}\ \bibnamefont {{Vincenti}}}, \bibinfo {author}
  {\bibfnamefont {G.}~\bibnamefont {{Leo}}},\ and\ \bibinfo {author}
  {\bibfnamefont {C.}~\bibnamefont {{De Angelis}}},\ }\href
  {https://doi.org/10.1109/JPHOT.2020.2988502} {\bibfield  {journal} {\bibinfo
  {journal} {IEEE Photonics Journal}\ }\textbf {\bibinfo {volume} {12}},\
  \bibinfo {pages} {1} (\bibinfo {year} {2020})}\BibitemShut {NoStop}%
\bibitem [{\citenamefont {Marino}\ \emph
  {et~al.}(2019{\natexlab{a}})\citenamefont {Marino}, \citenamefont {Gigli},
  \citenamefont {Rocco}, \citenamefont {Lema\^itre}, \citenamefont {Favero},
  \citenamefont {De~Angelis},\ and\ \citenamefont {Leo}}]{GM2019_2ndHarm}%
  \BibitemOpen
  \bibfield  {author} {\bibinfo {author} {\bibfnamefont {G.}~\bibnamefont
  {Marino}}, \bibinfo {author} {\bibfnamefont {C.}~\bibnamefont {Gigli}},
  \bibinfo {author} {\bibfnamefont {D.}~\bibnamefont {Rocco}}, \bibinfo
  {author} {\bibfnamefont {A.}~\bibnamefont {Lema\^itre}}, \bibinfo {author}
  {\bibfnamefont {I.}~\bibnamefont {Favero}}, \bibinfo {author} {\bibfnamefont
  {C.}~\bibnamefont {De~Angelis}},\ and\ \bibinfo {author} {\bibfnamefont
  {G.}~\bibnamefont {Leo}},\ }\href
  {https://doi.org/10.1021/acsphotonics.9b00110} {\bibfield  {journal}
  {\bibinfo  {journal} {ACS Photonics}\ }\textbf {\bibinfo {volume} {6}},\
  \bibinfo {pages} {1226} (\bibinfo {year} {2019}{\natexlab{a}})},\ \Eprint
  {https://arxiv.org/abs/https://doi.org/10.1021/acsphotonics.9b00110}
  {https://doi.org/10.1021/acsphotonics.9b00110} \BibitemShut {NoStop}%
\bibitem [{\citenamefont {Marino}\ \emph
  {et~al.}(2019{\natexlab{b}})\citenamefont {Marino}, \citenamefont {Solntsev},
  \citenamefont {Xu}, \citenamefont {Gili}, \citenamefont {Carletti},
  \citenamefont {Poddubny}, \citenamefont {Rahmani}, \citenamefont {Smirnova},
  \citenamefont {Chen}, \citenamefont {Lema\^{i}tre}, \citenamefont {Zhang},
  \citenamefont {Zayats}, \citenamefont {Angelis}, \citenamefont {Leo},
  \citenamefont {Sukhorukov},\ and\ \citenamefont {Neshev}}]{Marino:19}%
  \BibitemOpen
  \bibfield  {author} {\bibinfo {author} {\bibfnamefont {G.}~\bibnamefont
  {Marino}}, \bibinfo {author} {\bibfnamefont {A.~S.}\ \bibnamefont
  {Solntsev}}, \bibinfo {author} {\bibfnamefont {L.}~\bibnamefont {Xu}},
  \bibinfo {author} {\bibfnamefont {V.~F.}\ \bibnamefont {Gili}}, \bibinfo
  {author} {\bibfnamefont {L.}~\bibnamefont {Carletti}}, \bibinfo {author}
  {\bibfnamefont {A.~N.}\ \bibnamefont {Poddubny}}, \bibinfo {author}
  {\bibfnamefont {M.}~\bibnamefont {Rahmani}}, \bibinfo {author} {\bibfnamefont
  {D.~A.}\ \bibnamefont {Smirnova}}, \bibinfo {author} {\bibfnamefont
  {H.}~\bibnamefont {Chen}}, \bibinfo {author} {\bibfnamefont {A.}~\bibnamefont
  {Lema\^{i}tre}}, \bibinfo {author} {\bibfnamefont {G.}~\bibnamefont {Zhang}},
  \bibinfo {author} {\bibfnamefont {A.~V.}\ \bibnamefont {Zayats}}, \bibinfo
  {author} {\bibfnamefont {C.~D.}\ \bibnamefont {Angelis}}, \bibinfo {author}
  {\bibfnamefont {G.}~\bibnamefont {Leo}}, \bibinfo {author} {\bibfnamefont
  {A.~A.}\ \bibnamefont {Sukhorukov}},\ and\ \bibinfo {author} {\bibfnamefont
  {D.~N.}\ \bibnamefont {Neshev}},\ }\href
  {https://doi.org/10.1364/OPTICA.6.001416} {\bibfield  {journal} {\bibinfo
  {journal} {Optica}\ }\textbf {\bibinfo {volume} {6}},\ \bibinfo {pages}
  {1416} (\bibinfo {year} {2019}{\natexlab{b}})}\BibitemShut {NoStop}%
\bibitem [{\citenamefont {Carletti}\ \emph {et~al.}(2015)\citenamefont
  {Carletti}, \citenamefont {Locatelli}, \citenamefont {Stepanenko},
  \citenamefont {Leo},\ and\ \citenamefont {Angelis}}]{Carletti2015_OSA}%
  \BibitemOpen
  \bibfield  {author} {\bibinfo {author} {\bibfnamefont {L.}~\bibnamefont
  {Carletti}}, \bibinfo {author} {\bibfnamefont {A.}~\bibnamefont {Locatelli}},
  \bibinfo {author} {\bibfnamefont {O.}~\bibnamefont {Stepanenko}}, \bibinfo
  {author} {\bibfnamefont {G.}~\bibnamefont {Leo}},\ and\ \bibinfo {author}
  {\bibfnamefont {C.~D.}\ \bibnamefont {Angelis}},\ }\href
  {https://doi.org/10.1364/oe.23.026544} {\bibfield  {journal} {\bibinfo
  {journal} {Optics Express}\ }\textbf {\bibinfo {volume} {23}},\ \bibinfo
  {pages} {26544} (\bibinfo {year} {2015})}\BibitemShut {NoStop}%
\bibitem [{\citenamefont {Gili}\ \emph {et~al.}(2016)\citenamefont {Gili},
  \citenamefont {Carletti}, \citenamefont {Locatelli}, \citenamefont {Rocco},
  \citenamefont {Finazzi}, \citenamefont {Ghirardini}, \citenamefont {Favero},
  \citenamefont {Gomez}, \citenamefont {Lema\^{i}tre}, \citenamefont
  {Celebrano}, \citenamefont {Angelis},\ and\ \citenamefont {Leo}}]{Gili:16}%
  \BibitemOpen
  \bibfield  {author} {\bibinfo {author} {\bibfnamefont {V.~F.}\ \bibnamefont
  {Gili}}, \bibinfo {author} {\bibfnamefont {L.}~\bibnamefont {Carletti}},
  \bibinfo {author} {\bibfnamefont {A.}~\bibnamefont {Locatelli}}, \bibinfo
  {author} {\bibfnamefont {D.}~\bibnamefont {Rocco}}, \bibinfo {author}
  {\bibfnamefont {M.}~\bibnamefont {Finazzi}}, \bibinfo {author} {\bibfnamefont
  {L.}~\bibnamefont {Ghirardini}}, \bibinfo {author} {\bibfnamefont
  {I.}~\bibnamefont {Favero}}, \bibinfo {author} {\bibfnamefont
  {C.}~\bibnamefont {Gomez}}, \bibinfo {author} {\bibfnamefont
  {A.}~\bibnamefont {Lema\^{i}tre}}, \bibinfo {author} {\bibfnamefont
  {M.}~\bibnamefont {Celebrano}}, \bibinfo {author} {\bibfnamefont {C.~D.}\
  \bibnamefont {Angelis}},\ and\ \bibinfo {author} {\bibfnamefont
  {G.}~\bibnamefont {Leo}},\ }\href {https://doi.org/10.1364/OE.24.015965}
  {\bibfield  {journal} {\bibinfo  {journal} {Opt. Express}\ }\textbf {\bibinfo
  {volume} {24}},\ \bibinfo {pages} {15965} (\bibinfo {year}
  {2016})}\BibitemShut {NoStop}%
\bibitem [{\citenamefont {St-Jean}\ \emph {et~al.}(2017)\citenamefont
  {St-Jean}, \citenamefont {Goblot}, \citenamefont {Galopin}, \citenamefont
  {Lema{\^i}tre}, \citenamefont {Ozawa}, \citenamefont {Le~Gratiet},
  \citenamefont {Sagnes}, \citenamefont {Bloch},\ and\ \citenamefont
  {Amo}}]{St-Jean2017}%
  \BibitemOpen
  \bibfield  {author} {\bibinfo {author} {\bibfnamefont {P.}~\bibnamefont
  {St-Jean}}, \bibinfo {author} {\bibfnamefont {V.}~\bibnamefont {Goblot}},
  \bibinfo {author} {\bibfnamefont {E.}~\bibnamefont {Galopin}}, \bibinfo
  {author} {\bibfnamefont {A.}~\bibnamefont {Lema{\^i}tre}}, \bibinfo {author}
  {\bibfnamefont {T.}~\bibnamefont {Ozawa}}, \bibinfo {author} {\bibfnamefont
  {L.}~\bibnamefont {Le~Gratiet}}, \bibinfo {author} {\bibfnamefont
  {I.}~\bibnamefont {Sagnes}}, \bibinfo {author} {\bibfnamefont
  {J.}~\bibnamefont {Bloch}},\ and\ \bibinfo {author} {\bibfnamefont
  {A.}~\bibnamefont {Amo}},\ }\href {https://doi.org/10.1038/s41566-017-0006-2}
  {\bibfield  {journal} {\bibinfo  {journal} {Nature Photonics}\ }\textbf
  {\bibinfo {volume} {11}},\ \bibinfo {pages} {651} (\bibinfo {year}
  {2017})}\BibitemShut {NoStop}%
\bibitem [{\citenamefont {Armani}\ \emph {et~al.}(2003)\citenamefont {Armani},
  \citenamefont {Kippenberg}, \citenamefont {Spillane},\ and\ \citenamefont
  {Vahala}}]{Armani2003}%
  \BibitemOpen
  \bibfield  {author} {\bibinfo {author} {\bibfnamefont {D.~K.}\ \bibnamefont
  {Armani}}, \bibinfo {author} {\bibfnamefont {T.~J.}\ \bibnamefont
  {Kippenberg}}, \bibinfo {author} {\bibfnamefont {S.~M.}\ \bibnamefont
  {Spillane}},\ and\ \bibinfo {author} {\bibfnamefont {K.~J.}\ \bibnamefont
  {Vahala}},\ }\href {https://doi.org/10.1038/nature01371} {\bibfield
  {journal} {\bibinfo  {journal} {Nature}\ }\textbf {\bibinfo {volume} {421}},\
  \bibinfo {pages} {925} (\bibinfo {year} {2003})}\BibitemShut {NoStop}%
\bibitem [{\citenamefont {Baker}\ \emph {et~al.}(2014)\citenamefont {Baker},
  \citenamefont {Hease}, \citenamefont {Nguyen}, \citenamefont {Andronico},
  \citenamefont {Ducci}, \citenamefont {Leo},\ and\ \citenamefont
  {Favero}}]{Baker:14}%
  \BibitemOpen
  \bibfield  {author} {\bibinfo {author} {\bibfnamefont {C.}~\bibnamefont
  {Baker}}, \bibinfo {author} {\bibfnamefont {W.}~\bibnamefont {Hease}},
  \bibinfo {author} {\bibfnamefont {D.-T.}\ \bibnamefont {Nguyen}}, \bibinfo
  {author} {\bibfnamefont {A.}~\bibnamefont {Andronico}}, \bibinfo {author}
  {\bibfnamefont {S.}~\bibnamefont {Ducci}}, \bibinfo {author} {\bibfnamefont
  {G.}~\bibnamefont {Leo}},\ and\ \bibinfo {author} {\bibfnamefont
  {I.}~\bibnamefont {Favero}},\ }\href {https://doi.org/10.1364/OE.22.014072}
  {\bibfield  {journal} {\bibinfo  {journal} {Opt. Express}\ }\textbf {\bibinfo
  {volume} {22}},\ \bibinfo {pages} {14072} (\bibinfo {year}
  {2014})}\BibitemShut {NoStop}%
\bibitem [{\citenamefont {Parrain}\ \emph {et~al.}(2015)\citenamefont
  {Parrain}, \citenamefont {Baker}, \citenamefont {Wang}, \citenamefont {Guha},
  \citenamefont {Santos}, \citenamefont {Lemaitre}, \citenamefont {Senellart},
  \citenamefont {Leo}, \citenamefont {Ducci},\ and\ \citenamefont
  {Favero}}]{Parrain_Baker_OtpExp2015}%
  \BibitemOpen
  \bibfield  {author} {\bibinfo {author} {\bibfnamefont {D.}~\bibnamefont
  {Parrain}}, \bibinfo {author} {\bibfnamefont {C.}~\bibnamefont {Baker}},
  \bibinfo {author} {\bibfnamefont {G.}~\bibnamefont {Wang}}, \bibinfo {author}
  {\bibfnamefont {B.}~\bibnamefont {Guha}}, \bibinfo {author} {\bibfnamefont
  {E.~G.}\ \bibnamefont {Santos}}, \bibinfo {author} {\bibfnamefont
  {A.}~\bibnamefont {Lemaitre}}, \bibinfo {author} {\bibfnamefont
  {P.}~\bibnamefont {Senellart}}, \bibinfo {author} {\bibfnamefont
  {G.}~\bibnamefont {Leo}}, \bibinfo {author} {\bibfnamefont {S.}~\bibnamefont
  {Ducci}},\ and\ \bibinfo {author} {\bibfnamefont {I.}~\bibnamefont
  {Favero}},\ }\href {https://doi.org/10.1364/OE.23.019656} {\bibfield
  {journal} {\bibinfo  {journal} {Opt. Express}\ }\textbf {\bibinfo {volume}
  {23}},\ \bibinfo {pages} {19656} (\bibinfo {year} {2015})}\BibitemShut
  {NoStop}%
\bibitem [{\citenamefont {Roland}\ \emph {et~al.}(2020)\citenamefont {Roland},
  \citenamefont {Borne}, \citenamefont {Ravaro}, \citenamefont {Oliveira},
  \citenamefont {Suffit}, \citenamefont {Filloux}, \citenamefont
  {Lema\^{i}tre}, \citenamefont {Favero},\ and\ \citenamefont
  {Leo}}]{Roland:20}%
  \BibitemOpen
  \bibfield  {author} {\bibinfo {author} {\bibfnamefont {I.}~\bibnamefont
  {Roland}}, \bibinfo {author} {\bibfnamefont {A.}~\bibnamefont {Borne}},
  \bibinfo {author} {\bibfnamefont {M.}~\bibnamefont {Ravaro}}, \bibinfo
  {author} {\bibfnamefont {R.~D.}\ \bibnamefont {Oliveira}}, \bibinfo {author}
  {\bibfnamefont {S.}~\bibnamefont {Suffit}}, \bibinfo {author} {\bibfnamefont
  {P.}~\bibnamefont {Filloux}}, \bibinfo {author} {\bibfnamefont
  {A.}~\bibnamefont {Lema\^{i}tre}}, \bibinfo {author} {\bibfnamefont
  {I.}~\bibnamefont {Favero}},\ and\ \bibinfo {author} {\bibfnamefont
  {G.}~\bibnamefont {Leo}},\ }\href {https://doi.org/10.1364/OL.392417}
  {\bibfield  {journal} {\bibinfo  {journal} {Opt. Lett.}\ }\textbf {\bibinfo
  {volume} {45}},\ \bibinfo {pages} {2878} (\bibinfo {year}
  {2020})}\BibitemShut {NoStop}%
\bibitem [{\citenamefont {Pichugin}\ and\ \citenamefont
  {Sadreev}(2019)}]{Pichugin2019}%
  \BibitemOpen
  \bibfield  {author} {\bibinfo {author} {\bibfnamefont {K.~N.}\ \bibnamefont
  {Pichugin}}\ and\ \bibinfo {author} {\bibfnamefont {A.~F.}\ \bibnamefont
  {Sadreev}},\ }\href {https://doi.org/10.1063/1.5094188} {\bibfield  {journal}
  {\bibinfo  {journal} {Journal of Applied Physics}\ }\textbf {\bibinfo
  {volume} {126}},\ \bibinfo {pages} {093105} (\bibinfo {year} {2019})},\
  \Eprint {https://arxiv.org/abs/https://doi.org/10.1063/1.5094188}
  {https://doi.org/10.1063/1.5094188} \BibitemShut {NoStop}%
\bibitem [{\citenamefont {Ciuti}\ and\ \citenamefont
  {Carusotto}(2006)}]{Ciuti2006Input}%
  \BibitemOpen
  \bibfield  {author} {\bibinfo {author} {\bibfnamefont {C.}~\bibnamefont
  {Ciuti}}\ and\ \bibinfo {author} {\bibfnamefont {I.}~\bibnamefont
  {Carusotto}},\ }\href {https://doi.org/10.1103/physreva.74.033811} {\bibfield
   {journal} {\bibinfo  {journal} {Physical Review A}\ }\textbf {\bibinfo
  {volume} {74}},\ \bibinfo {pages} {033811} (\bibinfo {year}
  {2006})}\BibitemShut {NoStop}%
\bibitem [{\citenamefont {Zhang}\ \emph {et~al.}(2012)\citenamefont {Zhang},
  \citenamefont {Kang}, \citenamefont {Zhao}, \citenamefont {Zhou},\ and\
  \citenamefont {Lippens}}]{Zhang_2012_coupling}%
  \BibitemOpen
  \bibfield  {author} {\bibinfo {author} {\bibfnamefont {F.}~\bibnamefont
  {Zhang}}, \bibinfo {author} {\bibfnamefont {L.}~\bibnamefont {Kang}},
  \bibinfo {author} {\bibfnamefont {Q.}~\bibnamefont {Zhao}}, \bibinfo {author}
  {\bibfnamefont {J.}~\bibnamefont {Zhou}},\ and\ \bibinfo {author}
  {\bibfnamefont {D.}~\bibnamefont {Lippens}},\ }\href
  {https://doi.org/10.1088/1367-2630/14/3/033031} {\bibfield  {journal}
  {\bibinfo  {journal} {New Journal of Physics}\ }\textbf {\bibinfo {volume}
  {14}},\ \bibinfo {pages} {033031} (\bibinfo {year} {2012})}\BibitemShut
  {NoStop}%
\bibitem [{\citenamefont {Vial}\ and\ \citenamefont
  {Hao}(2016)}]{Vial_2016_coupling}%
  \BibitemOpen
  \bibfield  {author} {\bibinfo {author} {\bibfnamefont {B.}~\bibnamefont
  {Vial}}\ and\ \bibinfo {author} {\bibfnamefont {Y.}~\bibnamefont {Hao}},\
  }\href {https://doi.org/10.1088/2040-8978/18/11/115004} {\bibfield  {journal}
  {\bibinfo  {journal} {Journal of Optics}\ }\textbf {\bibinfo {volume} {18}},\
  \bibinfo {pages} {115004} (\bibinfo {year} {2016})}\BibitemShut {NoStop}%
\bibitem [{\citenamefont {Li}\ \emph {et~al.}(2021)\citenamefont {Li},
  \citenamefont {Soret},\ and\ \citenamefont {Ciuti}}]{Li2021}%
  \BibitemOpen
  \bibfield  {author} {\bibinfo {author} {\bibfnamefont {Z.}~\bibnamefont
  {Li}}, \bibinfo {author} {\bibfnamefont {A.}~\bibnamefont {Soret}},\ and\
  \bibinfo {author} {\bibfnamefont {C.}~\bibnamefont {Ciuti}},\ }\href
  {https://doi.org/10.1103/PhysRevA.103.022616} {\bibfield  {journal} {\bibinfo
   {journal} {Phys. Rev. A}\ }\textbf {\bibinfo {volume} {103}},\ \bibinfo
  {pages} {022616} (\bibinfo {year} {2021})}\BibitemShut {NoStop}%
\bibitem [{\citenamefont {Smirnova}\ \emph {et~al.}(2020)\citenamefont
  {Smirnova}, \citenamefont {Leykam}, \citenamefont {Chong},\ and\
  \citenamefont {Kivshar}}]{Kivshar_TopoRev_2020}%
  \BibitemOpen
  \bibfield  {author} {\bibinfo {author} {\bibfnamefont {D.}~\bibnamefont
  {Smirnova}}, \bibinfo {author} {\bibfnamefont {D.}~\bibnamefont {Leykam}},
  \bibinfo {author} {\bibfnamefont {Y.}~\bibnamefont {Chong}},\ and\ \bibinfo
  {author} {\bibfnamefont {Y.}~\bibnamefont {Kivshar}},\ }\href
  {https://doi.org/10.1063/1.5142397} {\bibfield  {journal} {\bibinfo
  {journal} {Applied Physics Reviews}\ }\textbf {\bibinfo {volume} {7}},\
  \bibinfo {pages} {021306} (\bibinfo {year} {2020})},\ \Eprint
  {https://arxiv.org/abs/https://doi.org/10.1063/1.5142397}
  {https://doi.org/10.1063/1.5142397} \BibitemShut {NoStop}%
\bibitem [{\citenamefont {Kruk}\ \emph {et~al.}(2019)\citenamefont {Kruk},
  \citenamefont {Poddubny}, \citenamefont {Smirnova}, \citenamefont {Wang},
  \citenamefont {Slobozhanyuk}, \citenamefont {Shorokhov}, \citenamefont
  {Kravchenko}, \citenamefont {Luther-Davies},\ and\ \citenamefont
  {Kivshar}}]{Kruk_Nature_2019}%
  \BibitemOpen
  \bibfield  {author} {\bibinfo {author} {\bibfnamefont {S.}~\bibnamefont
  {Kruk}}, \bibinfo {author} {\bibfnamefont {A.}~\bibnamefont {Poddubny}},
  \bibinfo {author} {\bibfnamefont {D.}~\bibnamefont {Smirnova}}, \bibinfo
  {author} {\bibfnamefont {L.}~\bibnamefont {Wang}}, \bibinfo {author}
  {\bibfnamefont {A.}~\bibnamefont {Slobozhanyuk}}, \bibinfo {author}
  {\bibfnamefont {A.}~\bibnamefont {Shorokhov}}, \bibinfo {author}
  {\bibfnamefont {I.}~\bibnamefont {Kravchenko}}, \bibinfo {author}
  {\bibfnamefont {B.}~\bibnamefont {Luther-Davies}},\ and\ \bibinfo {author}
  {\bibfnamefont {Y.}~\bibnamefont {Kivshar}},\ }\href
  {https://doi.org/10.1038/s41565-018-0324-7} {\bibfield  {journal} {\bibinfo
  {journal} {Nature Nanotechnology}\ }\textbf {\bibinfo {volume} {14}},\
  \bibinfo {pages} {126} (\bibinfo {year} {2019})}\BibitemShut {NoStop}%
\end{thebibliography}%
\end{document}


\maketitle

\section{Maxwell's equations resolution for an ensemble of Aluminum Gallium Arsenide nanocylinders}

\begin{figure}[H]
\begin{center}
\includegraphics[width=0.5\textwidth]{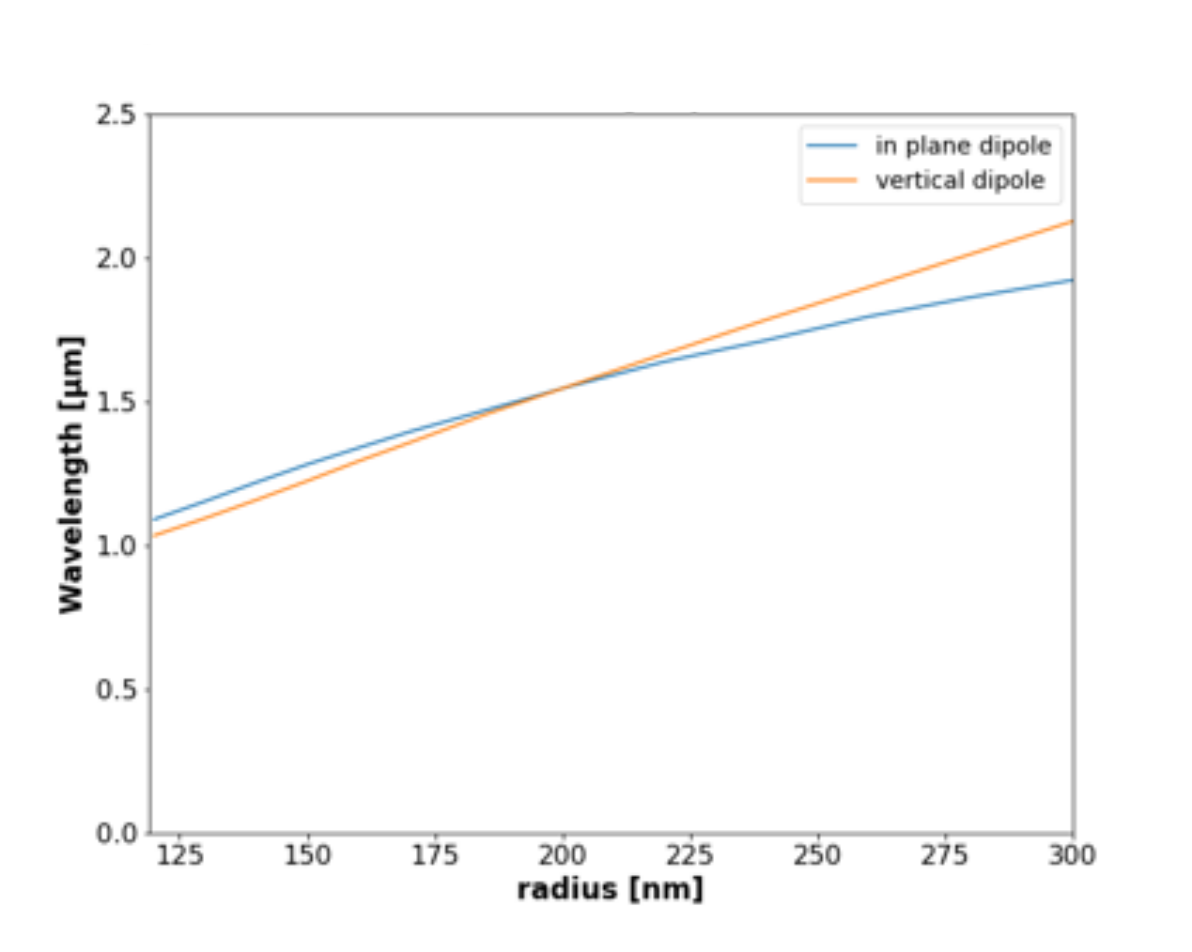}
\caption{Effect of nanocylinder radius on the real part of the eigenvalues of the magnetic dipolar modes for a 400~nm-high single nanocylinder. Orange curve corresponds to MD$_{z}$, and blue curve MD$_{x}$ and MD$_{y}$.}
\label{SM_mono}
\end{center}
\end{figure}

The proof-of-concept system for the non-Hermitian analytical quantum formalism consists of an ensemble of Al$_{0.18}$Ga$_{0.82}$As nanocylinders, which have recently drawn a great deal of attention because of the potential of dielectric nanostuctures on low-index substrate in terms of light control, nonlinear optics and topological physics. They can be seen as nanoresonators, whose properties are individually tailored by their geometry (Fig. \ref{SM_mono}). We have purposely selected cylindrical geometry to better convey the main message of our study. Eigenmodes of cylindrical structures are close to spherical resonators' ones, which provides a basis of mode close to Mie resonances. For equal diameter and height, in-plane dipoles (MD$_{x}$ and MD$_{y}$) and normal dipole (MD$_{z}$) are degenerate in energy. For a fixed height, varying the radius of the cylinder shifts the eigenfrequencies of dipolar modes, and lifts their degeneracy. Furthermore, nanocylinders are relatively easy to implement with e-beam lithography and are experimentally very accessible. Consequently, they give a large flexibility and represent a convenient toy system for more advanced characterisations, designs and future experiments. In practice, Al$_{0.18}$Ga$_{0.82}$As nanoantennas are supported by a low optical index semiconductor substrate like Aluminum oxide ($n_{AlOx} = 1.6$). However, since such substrate only results in second-order correction to the nanocylinders eigenmodes, here we neglect it and suppose the nanostructures to be surrounded by air.

\begin{figure}[ht!]
\begin{center}
\includegraphics[width=1\textwidth]{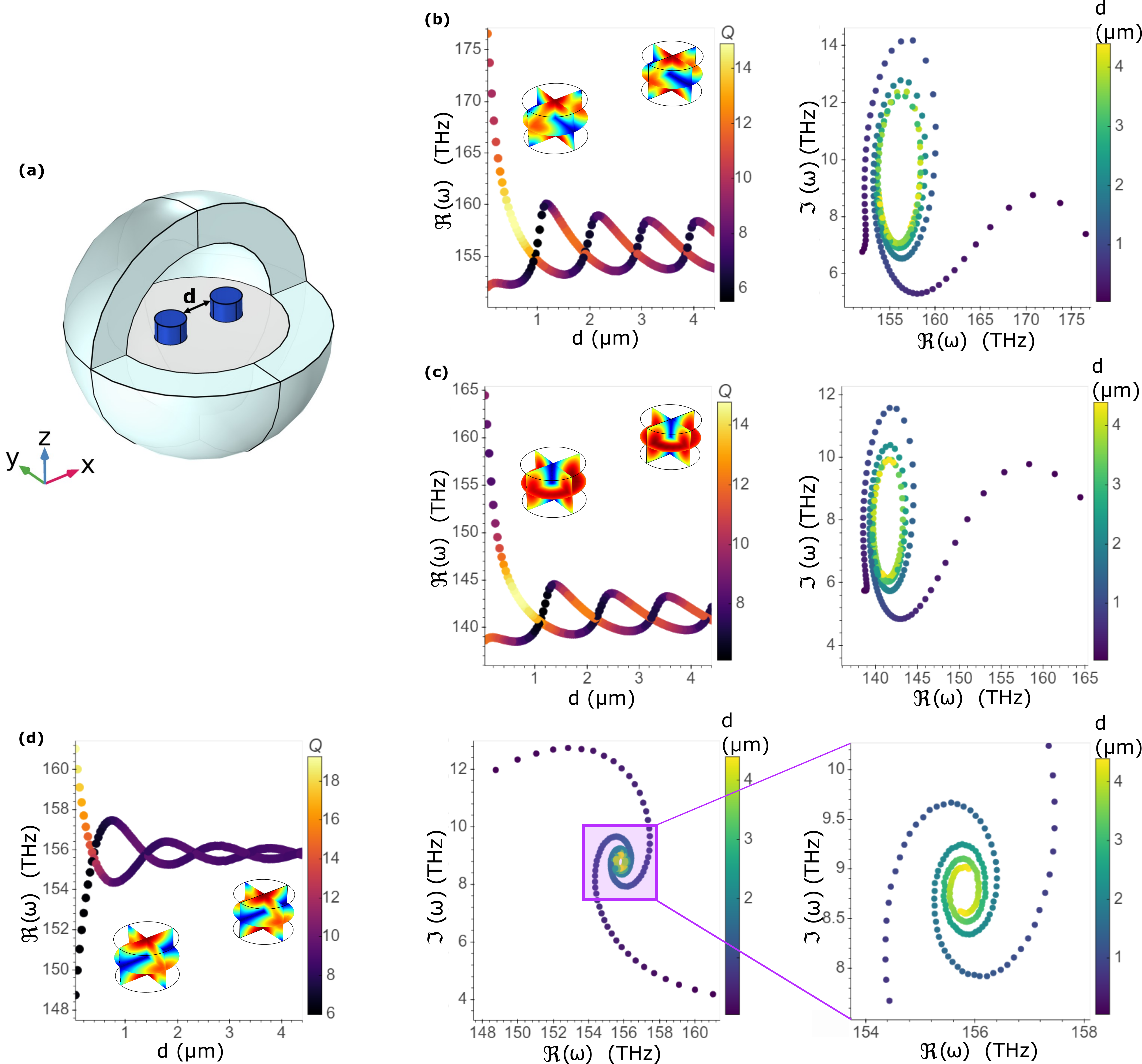}
\caption{FEM simulations in the case $N=2$, for MDs. \textbf{(a)} Schematic of the system: Al$_{0.18}$Ga$_{0.82}$As nanocylinders (dark blue); PML (light blue). \textbf{(b)}, \textbf{(c)} and \textbf{(d)} Left: real parts of the eigenfrequencies $\omega$ vs. gap for MD$_{y}$, MD$_{z}$ and MD$_{x}$, respectively. The color bar provides the quality factor $Q = \Re(\omega)/2\Im(\omega)$ and the insets are 3D plots of the electric field norm in the nanocylinders. Right: complex-plane representation of the same data, with the color bar providing the gap $d$.}
\label{SM2}
\end{center}
\end{figure}

To compute test data for the analytical non-Hermitian formalism (see Figs. 1 and 2 in main text, and Fig. \ref{SM2}), two algorithms were implemented. 

The former relies on near-field spatial correlations that enable us to track any given mode when the gap $d$ is varied. At each iteration, the electric field of each computed mode from FEM simulation is interpolated inside and in the close vicinity of each nanocylinder, and stored as an array of data. Spatial correlation coefficients are calculated by comparing the interpolated field from two successive iterations. With this method, by selecting the maximum of correlation coefficient one can reconstruct the evolution of a given mode when the gap is varied. Moreover, this algorithm can be initialized with the single element simulation: the modes of an $N$-element chain are then tracked by groups of $N$ hybrid modes stemming from a single-element mode.

The second algorithm relies on a direct identification of modes based on the evaluation of their complex field components in the whole integration space. For example, MD$_{z}$ has no $E_{z}$ component, and two equal $E_{y}$ and $E_{x}$ components. PML modes are discarded due to their higher field density in the PML layer. Each calculation of this algorithm is thus independent of the others. This procedure can be very versatile, assuming that the user can properly extract sorting condition from physics of Maxwell's equations. It proves more efficient with longer chains, for which eigenmode coupling to PML becomes more prominent in a finite calculation volume.


\section{Derivation of the quantum Langevin equation}
From the Hamiltonian (Eqs. (1-4) of the main text) and the commutation relations, the equation of motion for the continuum bath modes in the Heisenberg picture writes: 
\bea
\dfrac{\rmd \hat{\alpha_\eta}}{\rmd t}&=-i[\hat{\alpha_\eta},\hat{H}]\\&=-i\omega_\eta\hat{\alpha}_\eta-\sum_j g_{j\eta}^*\hat{a}_j,
\eea
which can be formally solved as
\bea
\hat{\alpha}_\eta(t)=e^{-i\omega_\eta(t-t_0)}\hat{\alpha}_\eta(t_0)-\sum_j g_{j\eta}^*\int_{t_0}^t \rmd t' e^{-i\omega	_\eta(t-t')}\hat{a}_j(t').
\eea
This leads to the dynamical equation for  $\hat{a}_j$:
\bea
\dfrac{\rmd \hat{a}_j}{\rmd t}=&-i[\hat{a}_j,\hat{H}_\text{sys}]+\int\rmd\eta~ g_{j\eta}\hat{\alpha}_\eta\\
=&-i[\hat{a}_j,\hat{H}_\text{sys}]+\int\rmd\eta~ g_{j\eta}e^{-i\omega_\eta(t-t_0)}\hat{\alpha}_\eta(t_0)\\&-\int\rmd\eta~\sum_k g_{j\eta}g_{k\eta}^*\int_{t_0}^t\rmd t' e^{-i\omega_\eta(t-t')}\hat{a}_k(t').
\eea
Defining 
\begin{align}
\hat{\alpha}_\eta^\text{in}&=\hat{\alpha}_\eta(t_0)e^{i\omega_\eta t_0},\\
\hat{F}_j(t)&=\int\rmd\eta~ g_{j\eta}e^{-i\omega_\eta t}\hat{\alpha}_\eta^\text{in},\\
\Gamma_{jk}(\tau)&=\Theta(\tau)\int\rmd\eta~ g_{j\eta}g_{k\eta}^* e^{-i\omega_\eta \tau},
\end{align}
where $\Theta(\tau)$ is the Heaviside step function. Under the limit $t_0\rightarrow -\infty$,
Eq. (S3) can be rewritten as 
\begin{equation}
\dfrac{\rmd \hat{a}_j}{\rmd t}=-i[\hat{a}_j,\hat{H}_\text{sys}]-\sum_k\int_{-\infty}^\infty \rmd t' \Gamma_{jk}(t-t')\hat{a}_k(t')+\hat{F}_j(t).
\end{equation}
Finally, the commutator can be calculated as 
\bea
\left[\hat{a}_j,\hat{H}_\text{sys}\right]&=\omega_0 \hat{a}_j+\sum_{j'}J(d_{jj'})\hat{a}_k,
\eea
which, together with the nearest-neighbour-coupling assumption, completes the derivation of the quantum Langevin equation (5) of the main text.

\section{Determining the coupling functions from the Maxwell equation solutions}

The eigenvalues of equation (10) in the main text are given by
\begin{equation}
\lambda_\pm(\omega,d)=\omega_0-\omega-\rmi\tilde{\Gamma}_\text{diag}(\omega,d)\pm(J(d)-\rmi\tilde{\Gamma}_\text{off}(\omega,d)),
\end{equation}
and therefore 
\begin{equation}\label{eq:dimerlambda}
    |\lambda_\pm(\omega,d)|^2=(\omega_0\pm J(d)-\omega)^2+(\tilde{\Gamma}_\text{diag}(\omega,d)\pm\tilde{\Gamma}_\text{off}(\omega,d))^2.
\end{equation}

\begin{figure}[ht]
\centering
\includegraphics[width=0.8\linewidth]{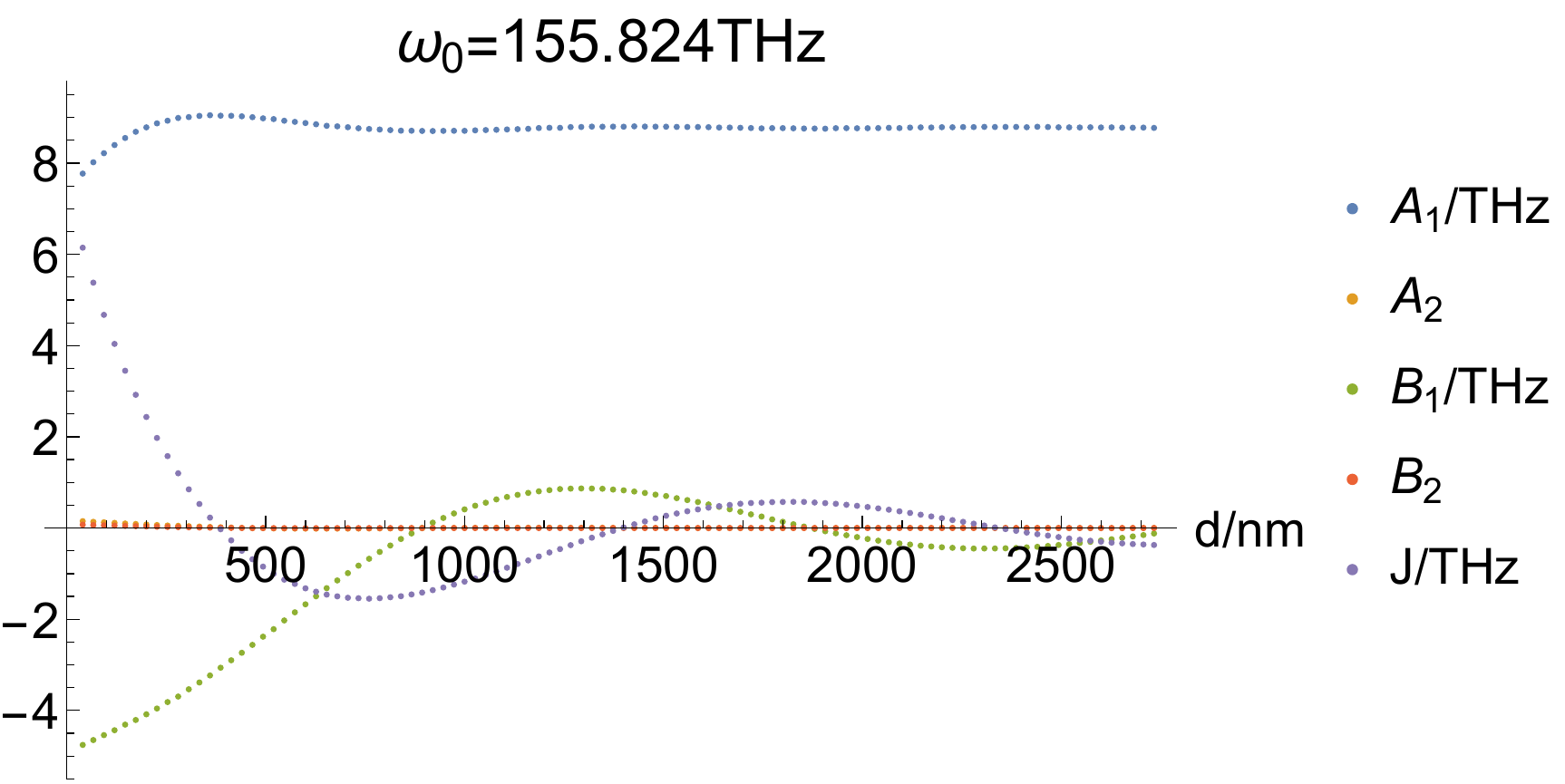}
\caption{Fit parameters for MD$_{x}$ obtained from our model.}\label{fig:fit}
\end{figure}

The simulation data of MD$_{x}$ for the $N=2$ case are shown in Fig. \ref{SM2} (d), with which we will demonstrate the determination of the coupling functions in our analytical model. For a simplified treatment, with no a priori knowledge on the frequency dependence of the reservoir functions, we expand them up to the first order in $\omega$:
\begin{align}
    \tilde{\Gamma}_\text{diag}(\omega,d)&=A_1(d)+A_2(d)(\omega-\omega_0),\\
    \tilde{\Gamma}_\text{off}(\omega,d)&=B_1(d)+B_2(d)(\omega-\omega_0).
\end{align}
This assumption makes Eq. (\ref{eq:dimerlambda}) quadratic in $\omega$, and the resonant frequencies $ \omega^\text{res}_\pm = \omega^\star_\pm - \rmi \gamma_\pm^\star$ can be expressed analytically in terms of our model parameters:
\begin{align}
    \omega^\star_+&=\frac{-A_1 A_2-A_1 B_2+A_2^2 \omega_0-A_2 B_1+2 A_2 B_2 \omega_0-B_1 B_2+B_2^2 \omega_0+J+\omega_0}{A_2^2+2 A_2 B_2+B_2^2+1},\\
    \omega^\star_-&=\frac{-A_1 A_2+A_1 B_2+A_2^2 \omega_0+A_2 B_1-2 A_2 B_2 \omega_0-B_1 B_2+B_2^2 \omega_0-J+\omega_0}{A_2^2-2 A_2 B_2+B_2^2+1},\\
    \gamma^\star_+&=A_1+B_1+(A_2+B_2)(\omega^\star_+-\omega_0),\\
    \gamma^\star_-&=A_1-B_1+(A_2-B_2)(\omega^\star_--\omega_0),
\end{align}
Solving this for each value of the gap $d$, we obtain a possible set of fit parameters plotted in Fig. \ref{fig:fit} as function of the gap, resulting in the same curves as in Fig. \ref{SM2} (d). 
While both sets $\{A_1,A_2,B_1,B_2,J\}$ and $\{ A_1,A_2,-B_1,-B_2,-J \}$ give the correct resonant frequencies $\omega_{\pm}^\mathrm{res}$, one of the solutions should be discarded in order to let the mode parity match. In particular, the FEM solutions can be identified as either "bonding" (non-zero electric field at midpoint between nanocylinders) or "antibonding" (zero electric field at midpoint) for each gap value. We then choose the solution set such that the eigenvector associated to an antibonding (resp. bonding) mode is proportional to (-1,1) (resp. (1,1)). This allows us to determine the fitted functions with no sign ambiguity, as shown in Fig.~\ref{fig:fit}. 

\section{Effectiveness of the analytical model}
\begin{figure}[ht]
    \centering
    \includegraphics[width=0.7\linewidth]{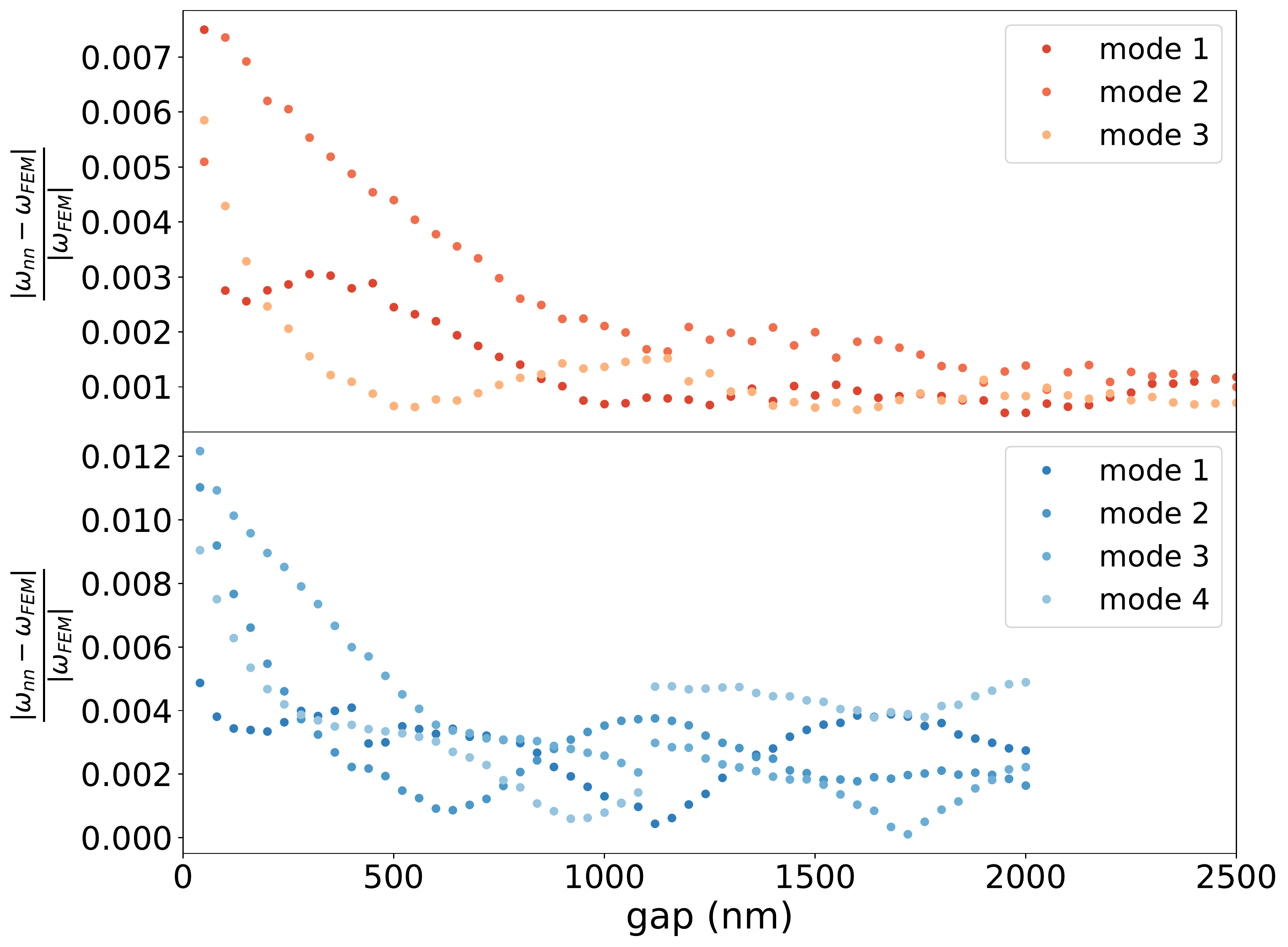}
    \caption{Relative error of the analytical model's prediction with only nearest-neighbour coupling as a function of the gap for the $N=3$ (top panel) and $N=4$ (bottom panel) systems.}
    \label{fig:err_nn}
\end{figure}

\begin{figure}[ht]
    \centering
    \includegraphics[width=0.7\linewidth]{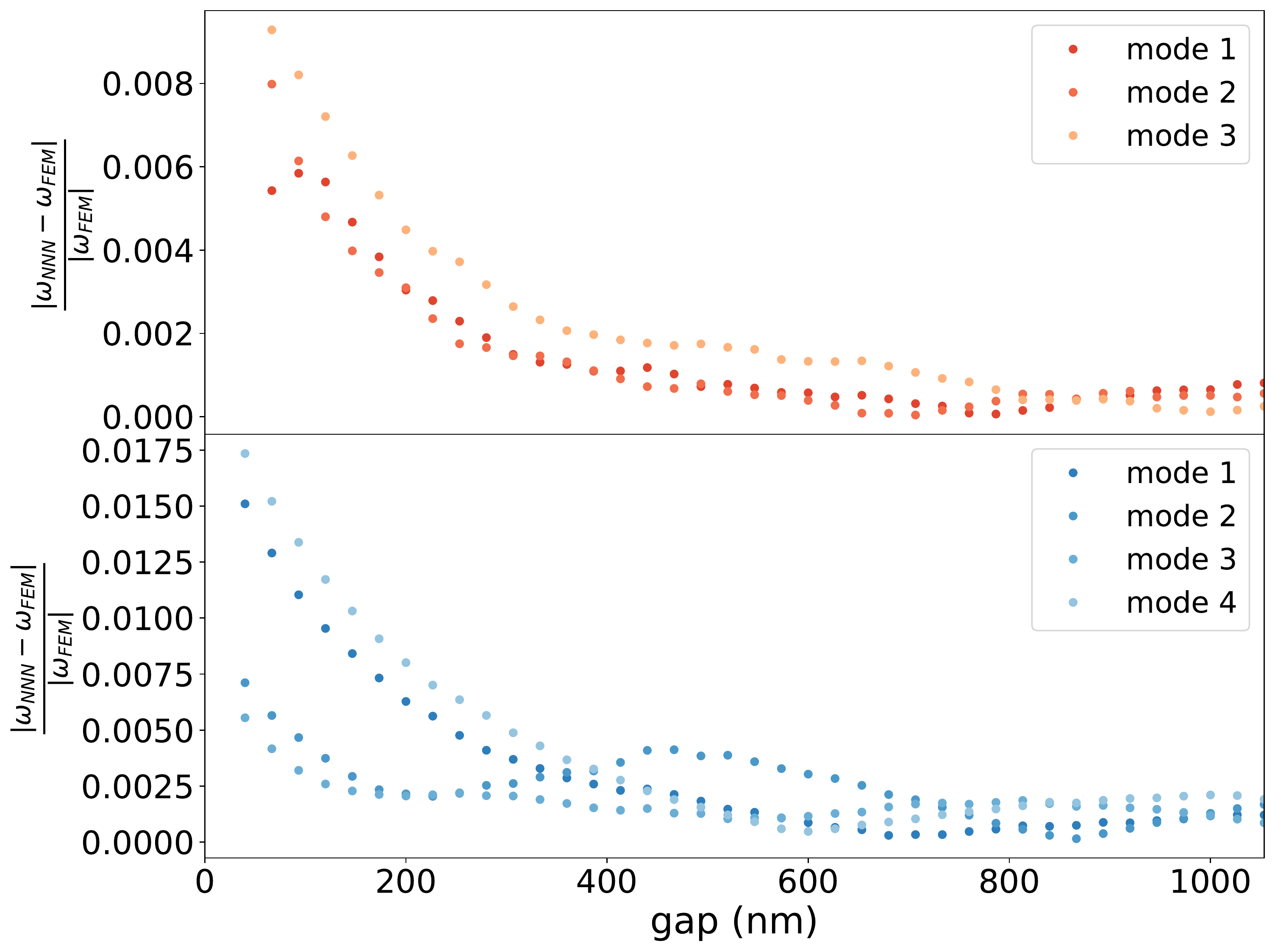}
    \caption{Relative error of the analytical model's prediction with simplified next-nearest-neighbour coupling as a function of the gap for the $N=3$ (top panel) and $N=4$ (bottom panel) systems.}
    \label{fig:err_NNN}
\end{figure}

To provide a measure of the efficacy of the approach, we quantify in Fig. \ref{fig:err_nn} and \ref{fig:err_NNN} the relative prediction error of the analytical model without and with the simplified next-nearest neighbour coupling for the cases $N=3$ and $N=4$, benchmarked against the FEM numerical solutions, which shows the good predictability of the analytical model.

